\documentclass[reprint,
superscriptaddress,
 amsmath,amssymb,
 aps
]{revtex4-1}

\usepackage{footmisc}
\usepackage{dcolumn}
\usepackage{bm}
\usepackage{xcolor} 
\usepackage{hyperref}       
\usepackage{url}            
\usepackage{booktabs}       
\usepackage{amsfonts}       
\usepackage{nicefrac}       
\usepackage{microtype}      
\usepackage{amsmath}
\usepackage{enumerate}   
\usepackage{mathtools}
\mathtoolsset{showonlyrefs=true}

\usepackage{graphicx} 
\usepackage{float}
\graphicspath{{Figures/}} 

\usepackage{xcolor}

\newcommand{\probtarget}[1]{p_{\text{target}}\left( #1 \right)}
\newcommand{\probbg}[1]{p_{\text{bg}}\left( #1 \right)}
\newcommand{\probunif}[1]{p_{\text{Unif}}\left( #1 \right)}
\newcommand{\probp}[1]{p\left( #1 \right)}
\newcommand{\probq}[1]{q\left( #1 \right)}

\newcommand{\sampleX}{\mathbf{x}}

\newcommand{\sampleU}{\mathbf{u}}
\newcommand{\probU}[1]{p_{\mathbf{u}}\left( #1 \right)}

\newcommand{\sampleZ}{\mathbf{z}}
\newcommand{\probqphi}[1]{q_\phi\left(#1 \right)}

\newcommand{\expectationOver}[2]{\mathbb{E}_{#1}\left[ #2 \right]}
\newcommand{\kmax}{k_\text{max}}
\newcommand{\kQ}[1]{k_{\text{Q}#1}}
\newcommand{\wQ}[1]{w_{\text{Q}#1}}

\newcommand{\comment}[1]{}

\DeclareMathOperator*{\argmax}{arg\,max}
\DeclareMathOperator*{\argmin}{arg\,min}

\begin{document}

\title{Exhaustive Neural Importance Sampling applied to Monte Carlo event generation}

\author{Sebastian Pina-Otey}
\email{pinas@aia.es}
\affiliation{Aplicaciones en Inform\'{a}tica Avanzada (AIA), Sant Cugat del Vall\`{e}s (Barcelona) 08172, Spain}
\affiliation{Institut de F\'{i}sica d\'{}Altes Energies (IFAE) - Barcelona Institute of Science and Technology (BIST), Bellaterra (Barcelona) 08193, Spain}

\author{Federico S\'{a}nchez}
\affiliation{University of Geneva, Section de Physique, DPNC, Geneva 1205, Switzerland}

\author{Thorsten Lux}
\affiliation{Institut de F\'{i}sica d\'{}Altes Energies (IFAE) - Barcelona Institute of Science and Technology (BIST), Bellaterra (Barcelona) 08193, Spain}

\author{Vicens Gaitan}
\affiliation{Aplicaciones en Inform\'{a}tica Avanzada (AIA), Sant Cugat del Vall\`{e}s (Barcelona) 08172, Spain}

\date{\today}

\begin{abstract}
The generation of accurate neutrino-nucleus cross-section models needed for neutrino oscillation experiments require simultaneously the description of many degrees of freedom and precise calculations to model nuclear responses. The detailed calculation of complete models makes the Monte Carlo generators slow and impractical. We present Exhaustive Neural Importance Sampling (ENIS), a method based on normalizing flows to find a suitable proposal density for rejection sampling automatically and efficiently, and discuss how this technique solves common issues of the rejection algorithm. 

\end{abstract}

\maketitle

\section{ Motivation }

In modern science and engineering disciplines, the generation of random samples from a probability density function to obtain data sets or compute expectation values has become an essential tool. These theoretical models can be described by a target probability density function $\probp{\sampleX}$. Ideally, to generate samples following $\probp{\sampleX}$, the inverse transformation method is used. To perform the inverse transformation, the cumulative probability has to be calculated and the inverse to this function has to be found. Numerical methods have to be applied to obtain the Monte Carlo (MC) samples when this is not feasible computationally.  This is especially true for high-dimensional spaces, where the integrals required to find such inverse transformation become analytically challenging.

A standard numerical method to obtain such data set is to perform a Markov Chain Monte Carlo (MCMC) algorithm \cite{10.1093/biomet/57.1.97}, which provides good results for expected value calculations. Compared to other methods, it has the advantage that, in general, it requires very little calibration, and high dimensions can be broken down into conditional smaller dimension densities \cite{4767596}. However, the MCMC method produces samples that form a correlated sequence. Also, the convergence of the samples' chain to the target density cannot be guaranteed for all possible models.

Another standard algorithm to produce MC samples is the acceptance-rejection or simply rejection sampling \cite{casella,MARTINO20102981,Bishop2007PatternRA,vonNeumann1951}, which produces i.i.d. (independent and identically distributed) samples from the target density via an auxiliary proposal function. The proposal has to satisfy being a density which can both be sampled from and evaluated efficiently, as well as being as close to the target density as possible. The main disadvantages of the method are the following \cite{Robert2004}:
\begin{enumerate}
    \item Designing the proposal function close to a particular target density can be very costly in human time.
    \item If a generic proposal function is taken, such as a uniform distribution over the domain, the algorithm is usually very inefficient.
    \item The sampling efficiency decreases rapidly with the number of dimensions.
\end{enumerate}
Ideally, to avoid these inconveniences, one would like to have a method to find a proposal function that adapts to a given target density automatically. This would solve simultaneously the human time cost as well as the inefficiency of generic proposal densities. 

An approach of the usage of normalizing flows to find a suitable proposal for a given target density has been suggested previously as Neural Importance Sampling (NIS) \cite{Mller2018NeuralIS}, focused on the integration of functions via importance sampling \cite{10.2307/166789}. Normalizing flows provide an expressive family of parametrized density functions $\probqphi{\sampleX}$ through neural networks, by defining a differentiable and invertible transformation from a base distribution to a target distribution, allowing to evaluate and sample from complex distributions by transforming simpler ones. The concept of integrating via importance sampling with normalizing flows for High Energy Physics (HEP) has been explored in other works to obtain top-quark pair production and gluon-induced multi-jet production \cite{Bothmann:2020ywa} or to simulate collider experimental observables for the Large Hadron Collider \cite{PhysRevD.101.076002}.

In this work we further explore the possibility of utilizing normalizing flows to find a proposal function for a given target density to perform rejection sampling for MC samples, and analyze its viability through the following points:
\begin{itemize}
    \item We discuss the importance of adding an additional density (background) to the target one to assure the coverage of the whole phase space when performing rejection sampling.
    \item We define a two-phase training scheme for the normalizing flow to boost initial inefficiency  in the optimization when adjusting the initialized random density towards the target one. 
    \item We measure the performance of the method and argue for relaxing the rejection sampling constant factor $k$ to improve largely the efficiency of acceptance while quantifying the error committed in doing this approximation via the concept of coverage.
\end{itemize}
Considering the proposed algorithm covers the whole domain of interest by modifying NIS with the background density, we denote this method by Exhaustive Neural Importance Sampling (ENIS).

We apply the above algorithm to a HEP problem, in the form of the charged current quasi-elastic (CCQE) cross-section for anti-neutrinos interactions with nuclei, performing in-depth analysis and discussion of the efficiency of the method. 
Neutrino-nucleus cross-section modeling is one of the main sources of systematic uncertainties in neutrino oscillations measurements \cite{Alvarez-Ruso:2017oui,Abe:2019vii,Acero:2019ksn}. Cross-section models are either analytically simple but describe the experimental data poorly or involving complex numerical computations, normally related to the description of the nucleus, that imposes limitations in their MC implementation. New tendencies in the field also call for a fully exclusive description of the interaction adding complexity to the calculations. The analytical model utilized in this paper is simple, but it is a realistic one and a good reference to demonstrate the capabilities of the proposed method to generate  neutrino-nucleus cross-sections efficiently. We will show that ENIS opens the possibility to incorporate efficiently complex theoretical models in the existing MC models enhancing the physics reach of running and future neutrino oscillation experiments. 

ENIS algorithm may be used beyond the scope of neutrino physics. Further applications to be evaluated in detail in the future are particle/nuclear physics experiments, detector responses for medical physics, engineering studies or theoretical modelling. In general, it could be applied to any Monte Carlo simulation that is limited by the algorithm's  speed, such as for importance sampling to provide fast Monte Carlo with sufficient accuracy (i.e. fast detector simulation, design studies, minimum bias background simulations, etc.). Additionally, the technique may help model developers extract expected values from their theoretical predictions in realistic conditions by including simple  detector effects in models, such as  effects of detector acceptance cuts, impact of model degrees of freedom on the predictions or uncertainty propagation.

\section{Framework}
\label{Sec:Background}

In this Section we will describe background and framework needed for the rest of the paper. Sec.~\ref{Sec:CCQEXsect} explains the physical model of charged current quasi-elastic neutrino interaction we will apply ENIS to in Sec.~\ref{Sec:MCCCQE}. As a summary and to introduce our notation, Sec.~\ref{Sec:RejectionSampling} overviews the rejection sampling algorithm. Finally, in Sec.~\ref{Sec:NSFTheory} we make a modest introduction to normalizing flows, focusing on the implementation of Neural Spline Flows.

\subsection{Model of Charged Current Quasi-Elastic Anti-Neutrinos Interactions with Nuclei}
\label{Sec:CCQEXsect}

 The Charged Current Quasi-elastic (CCQE) is a basic model of neutrino interactions that might be expressed in simple formulae. The CCQE model has many advantages during this exploratory work, as it can be implemented in a simple software function, while at the same time it is also a realistic environment to understand the implications of modeling cross-sections with the proposed methodology. The selected model to describe CCQE is the well established Smith-Monith \cite{Smith:1972xh}. The nucleon momentum distribution follows a Relativistic Fermi Gas (non-interacting nucleons in a nuclear potential well) with a $0.225$~GeV/c Fermi level. The model includes the Pauli blocking effect, preventing the creation of final state nucleons below the nucleus Fermi level. The model can be applied both to neutrino and antineutrino interactions. Antineutrinos are selected for this study due to the vector axial current cancellation imposing more stringent conditions at the edges of the kinematic phase space. The model includes the following degrees of freedom generated by the Monte Carlo model: the neutrino energy, the $\mu^{\pm}$ momentum and angle, and the target nucleon Fermi momentum.  Contrary to other MC implementations, the neutrino energy is not a fixed input value but it is generated by the algorithm to add complexity to the calculations and to check the capabilities of the calculations to reproduce the cross-section as a function of the neutrino energy. The implementation of this model for fixed energy value is also possible. The basic kinematic distributions obtained with this model will be discussed in Sec.~\ref{Sec:MCCCQE}.
 
 \subsection{Rejection sampling}
\label{Sec:RejectionSampling}

Rejection sampling is a well known technique \cite{casella,MARTINO20102981,Bishop2007PatternRA,vonNeumann1951,Robert2004} to obtain MC samples from a target density $\probp{\sampleX}$ which can be evaluated (up to a constant), but cannot be sampled from through the inverse transform. It relies on an auxiliary proposal function $\probq{\sampleX}$, from which one should be able to sample from and evaluate efficiently. A constant $k>0$ is introduced which has to satisfy that
\begin{equation}
    k \cdot \probq{\sampleX}\geq \probp{\sampleX}\;\; \forall \;\sampleX:\probp{\sampleX}>0. \label{Eq:rejection}
\end{equation}
The resulting function $k \cdot \probq{\sampleX}$ is called the comparison function. 

The procedure to sample from the target density is then the following:
\begin{enumerate}
    \item A sample $\sampleX$ is generated following $\probq{\sampleX}$, $\sampleX\sim\probq{\sampleX}$.
    \item A random number $u$ is generated uniformly in the range $[0,k \cdot \probq{\sampleX}]$, $u\sim \text{Unif}(0,k \cdot \probq{\sampleX})$.
    \item If $u$ fulfills the condition $u\leq \probp{\sampleX}$, the sample is accepted; otherwise, it is rejected. 
\end{enumerate}

Additionally, if $\probp{\sampleX}$ is normalized, the probability that a sample is accepted is proportional to $p_{\text{acc}}\propto 1/k$, i.e., $k$ gives an intuition of the number of tries until we obtain an accepted sample.
 
\subsection{Neural density estimation using Neural Spline Flows}\label{Sec:NSFTheory}

A family of density functions $\probqphi{\sampleX}$ over the real D-dimensional space $\mathbb{R}^D$ parametrized by $\phi$ satisfies that $\probqphi{\sampleX} \geq 0$ for all $\sampleX,\phi$, and that $\int \probqphi{\sampleX} \; d\sampleX = 1$ for all $\phi$. Normalizing flows are a mechanism of constructing such flexible probability density families $\probqphi{\sampleX}$ for continuous random variables $\sampleX\in\mathbb{R}^D$.  A comprehensive review on the topic can be found in \cite{Papamakarios2019NormalizingFF}, from which a brief summary will be shown in this Section on how normalizing flows are defined, and how the parameters $\phi$ are obtained, together with a specific implementation, the Neural Spline Flows (NSF) \cite{Durkan2019NeuralSF}.

Consider a random variable $\sampleU$ defined over $\mathbb{R}^D$, with known probability density $\probU{\sampleU}$. A normalizing flow characterizes itself by a transformation $T$ from another density $\probp{\sampleX}$ of a random variable $\sampleX$ defined also over $\mathbb{R}^D$, the target density, to this known density, via
\begin{align}\label{Eq:transformation}
    \sampleU = T(\sampleX)\text{, with }\sampleX\sim\probp{\sampleX}.
\end{align}
The density $\probU{\sampleU}$ is known as base density, and has to satisfy that it is easy to evaluate (e.g., a multivariate $D$-dimensional normal distribution, as will be chosen through this work, or a uniform distribution in dimension $D$). The transformation $T$ has to be invertible, and both $T$ and $T^{-1}$ have to be differentiable, i.e., $T$ defines a diffeomorphism over $\mathbb{R}^D$.

This allows us to evaluate the target density by evaluating the base density using the change of variables for density functions,
\begin{align}
    \probp{\sampleX} = \probU{T(\sampleX)} |\det J_T(\sampleX)|,
\end{align}
where the Jacobian $J_T(\sampleX)$ is a $D\times D$ matrix of the partial derivatives of the transformation $T$:
\begin{align}
    J_T(\sampleX) = \left[\begin{array}{ccc}
         \frac{\partial T_1}{\partial x_1} & \cdots & \frac{\partial T_1}{\partial x_D}  \\
         \vdots & \ddots & \vdots \\
         \frac{\partial T_D}{\partial x_1} & \cdots & \frac{\partial T_D}{\partial x_D}
    \end{array}\right].
\end{align}

The transformation $T$ in a normalizing flow is defined partially through a neural network with parameters $\phi$, as will be described below, defining a density
\begin{align}\label{Eq:changeVariables}
    \probqphi{\sampleX} = \probU{T_\phi(\sampleX)} |\det J_{T_\phi}(\sampleX)|.
\end{align}
The subindex of $T_\phi$ will be omitted in the following, simply denoting the transformation of the neural network by $T$.

If the transformation is flexible enough, the flow could be used to evaluate any continuous density in $\mathbb{R}
^D$. In practice, however, the property that the composition of diffeomorphisms is a diffeomorphism is used, allowing to construct a complex transformation via composition of simpler transformations. Consider the transformation $T$ as a composition of simpler $T_k$ transformations:
\begin{align}
    T=T_K\circ \cdots \circ T_1.
\end{align}
Assuming $\sampleZ_0=\sampleX$ and $\sampleZ_K = \sampleU$, the forward evaluation and Jacobian are
\begin{align}
    \sampleZ_k &= T_k(\sampleZ_{k-1}),\; k=1:K,\\
    |J_T(\sampleX)| &= \left|\prod_{k=1}^K J_{T_k}(\sampleZ_{k-1})  \right|.
\end{align}
These two computations (plus their inverse) are the building blocks of a normalizing flow \cite{Rezende2015VariationalIW}. Hence, to make a transformation efficient, both operations have to be efficient. From now on, we will focus on a simple transformation $\sampleU=T(\sampleX)$, since constructing a flow from it is simply applying compositions.

To define a transformation satisfying both operations to be efficient, the transformation is broken down into autoregressive one-dimensional ones for each dimension of $\mathbb{R}^D$:
\begin{align}
u_i = \tau(x_i;\mathbf{h}_i)\text{ with } \mathbf{h}_i = c_i(\sampleX_{<i};\phi),
\end{align}
where $u_i$ is the $i$-th component of $\sampleU$ and $x_i$ the $i$-th of $\sampleX$.  $\tau$ is the transformer, which is a one-dimensional diffeomorphism with respect to $x_i$ with parameters $\mathbf{h}_i$. $c_i$ is the $i$-th conditioner, a neural network, which takes as input $\sampleX_{<i}=(x_1,x_2,\dots,x_{i-1})$, i.e., the previous components of $\sampleX$, and $\phi$ are the parameters of the neural network. The conditioner provides the parameters $\mathbf{h}_i$ of the $i$-th transformer of $x_i$ depending on the previous components $\sampleX_{<i}$, defining implicitly a conditional density over $x_i$ with respect to $\sampleX_{<i}$. The transformer is chosen to be a differentiable monotonic function, since then it satisfies the requirements to be a diffeomorphism. The transformer also satisfies that it makes the transformation easily computable in parallel and decomposing the transformation in one dimensional autoregressive transformers allows the computation of the Jacobian to be trivial, because of its triangular shape. To compute the parameter $\mathbf{h}_i$ of each transformer, one would need to process a neural network with input $\sampleX_{<i}$ for each component, a total of $D$ times. 

Masked autoregressive neural networks \cite{Germain2015MADEMA} enable us to compute all the conditional functions simultaneously in a single forward iteration of the neural network. This is done by masking out, with a binary matrix, the connections of the $\mathbf{h}_i$-th output with respect to all the components with index bigger or equal to $i$, $\geq i$, making it a function of the $<i$ components.

The transformer can be defined by any monotonic function, such as affine transformations \cite{Papamakarios2017MaskedAF}, monotonic neural networks \cite{Huang2018NeuralAF,Cao2019BlockNA,Wehenkel2019UnconstrainedMN}, sum-of-squares polynomials \cite{Jaini2019SumofSquaresPF} or monotonic splines \cite{Mller2018NeuralIS,Durkan2019CubicSplineF,Durkan2019NeuralSF}. In this work we will focus on a specific implementation of monotonic splines, the Neural Spline Flows.

In their work on Neural Spline Flows \cite{Durkan2019NeuralSF}, Durkan et al. advocate for utilizing monotonic rational-quadratic splines as  transformers $\tau$, which are easily differentiable, more flexible than previous attempts of using polynomials for these transformers, since their Taylor-series expansion is infinite, and are analytically invertible. 

Each monotonic rational-quadratic function in the splines is defined by a quotient of two quadratic polynomial. In particular, the splines map the interval $[-B,B]$ to $[-B,B]$, and outside of it the identity function is considered. The splines are parametrized following Gregory and Delbourgo \cite{Gregory1982PiecewiseRQ}, where $K$ different rational-quadratic functions are used, with boundaries set by the pair of coordinates $\{(x^{(k)},u^{(k)}\}_{k=0}^K$, known as knots of the spline and are the points where it passes through. Note that $(x^{(0)},u^{(0)}) = (-B,-B)$ and $(x^{(K)},u^{(K)}) = (B,B)$. Additionally, we need $K-1$ intermediate positive derivative values, since the boundary points derivatives are set to 1 to match the identity function. 

Having this in mind, the conditioner given by the neural network outputs a vector $\mathbf{h}=[\mathbf{h}^w,\mathbf{h}^h,\mathbf{h}^d]$ of dimension $(3\times K-1)$ for the transformer $\tau$, $c_i(\sampleX_{<i};\phi)=\mathbf{h}_i$. $\mathbf{h}^w$ and $\mathbf{h}^h$ give the width and height of the $K$ bins, while $\mathbf{h}^d$ is the positive derivative at the intermediate $(K-1)$ knots. 

Stacking up many of these transformations, a highly flexible neural density estimator, the NSF, can be build, which satisfies:
\begin{enumerate}
    \item It is easy to sample from $\probqphi{\sampleX}$ using the inverse transform $T^{-1}$ in Eq.~\eqref{Eq:transformation} by sampling $\sampleU\sim\probU{\sampleU}$.
    \item Eq.~\eqref{Eq:changeVariables} allows to evaluate the densities $\probqphi{\sampleX}$ of these samples when generating them in an efficient way.
\end{enumerate}
This density estimator will be the one utilized during this work.

\section{Methodology}
\label{Sec:Method}

With the framework introduced in Sec.~\ref{Sec:Background}, we are now in  a position to define the ENIS method and the different metrics we will use to measure its performance.

We start in Sec.~\ref{Sec:Optimization} by showing the objective function to be minimized by the NSF to adjust its proposal function $\probqphi{\sampleX}$ to the target density $\probp{\sampleX}$. Then, in Sec.~\ref{Sec:BackgrounNoise}, we discuss the importance of adding background noise to both ensure coverage of the whole phase space of $\probp{\sampleX}$ and to boost the initial phase of the training of ENIS. The exact training scheme is then shown in Sec.~\ref{Sec:TrainingScheme}, differentiating the warm-up phase from the iterative phase. Finally, in Sec.~\ref{Sec:MeasuringPerformance}, the performance metrics are introduced, explaining the concept of coverage and effective sample size when considering a more relaxed condition on the rejection constant $k$. 

\subsection{Optimizing the parameters of the NSF}
\label{Sec:Optimization}

Consider a target probability density function $\probp{\sampleX}$ which can be evaluated for all $\sampleX$ but from which we are unable to generate samples directly through an analytical inverse transform. If we could approximate this target density by our neural density estimator $\probqphi{\sampleX}$, then we could exactly sample from the target density using rejection sampling, since we can both sample and evaluate from $\probqphi{\sampleX}$.

To obtain the parameters $\phi$ of $\probqphi{\sampleX}$ given a density $\probp{\sampleX}$ which can be evaluated, we want to minimize the Kullback-Leibler divergence (KL-divergence) \cite{kullback1951} between both distributions, which is $\geq 0$ and only equal to zero if both distributions match:
\begin{align}\label{Eq:KLdivergence}
    D_{\text{KL}}\big(\probp{\sampleX}\| \probqphi{\sampleX}\big) = \int \probp{\sampleX} \log \left( \frac{\probp{\sampleX}}{\probqphi{\sampleX}}\right) \;d\sampleX.
\end{align}

When minimizing with respect to $\phi$, the KL-divergence is simplified to
\begin{align}
    &\argmin_\phi D_{\text{KL}}\big(\probp{\sampleX}\| \probqphi{\sampleX}\big)\\ 
    & \phantom{\argmin_\phi D_{\text{KL}}} = \argmin_\phi \int \probp{\sampleX} \log \left( \frac{\probp{\sampleX}}{\probqphi{\sampleX}}\right) d\sampleX\\
    & \phantom{\argmin_\phi D_{\text{KL}}}= \argmin_\phi -\int \probp{\sampleX} \log \probqphi{\sampleX} \;d\sampleX\\
    & \phantom{\argmin_\phi D_{\text{KL}}}= \argmax_\phi \int \probp{\sampleX} \log \probqphi{\sampleX} \;d\sampleX\\
    & \phantom{\argmin_\phi D_{\text{KL}}}=\argmax_\phi \expectationOver{\sampleX\sim\probp{\sampleX}}{\log \probqphi{\sampleX}}.\label{Eq:lossFunction}
\end{align}
The last expression could be approximated numerically if we could sample $\sampleX\sim\probp{\sampleX}$, since it corresponds to approximating an expected value of a function we can evaluate, $\log \probqphi{\sampleX}$, and is equal to maximizing the log-likelihood of these samples: 
\begin{align}\label{Eq:objective}
    L = \frac{1}{N} \sum_{i=1}^N \log(\probqphi{\sampleX_i})  \text{ with } \sampleX_i\sim \probp{\sampleX}.
\end{align}

M{\"u}ller et al. \cite{Mller2018NeuralIS} propose a solution for computing the gradient with respect to $\phi$ for this maximization problem. They suggest using importance sampling \cite{10.2307/166789} for this particular expected value:
\begin{align}\label{Eq:gradientImportance}
  \nabla_\phi \expectationOver{\sampleX\sim\probp{\sampleX}}{\log \probqphi{\sampleX}} &= \int \probp{\sampleX} \nabla_\phi \log \probqphi{\sampleX} \;d\sampleX\\
   &=\int \probqphi{\sampleX} \frac{\probp{\sampleX}}{\probqphi{\sampleX}} \nabla_\phi \log \probqphi{\sampleX} \;d\sampleX\\
   & =  \expectationOver{\sampleX\sim\probqphi{\sampleX}}{w(\sampleX) \nabla_\phi \log \probqphi{\sampleX}}\\
   &\approx \frac{1}{N} \sum_{i=1}^N w(\sampleX_i) \nabla_\phi \log \probqphi{\sampleX_i},\label{Eq:gradientImportanceApprox}
\end{align}
with $\sampleX_i\sim \probqphi{\sampleX}$ and the weights defined as $w(\sampleX) = \probp{\sampleX}/\probqphi{\sampleX}$. Notice how we only need to be able to evaluate $\probp{\sampleX}$ to compute this quantity. With this gradient, we are able to minimize the KL-divergence in Eq.~\eqref{Eq:KLdivergence} if the support of $\probqphi{\sampleX}$ (i.e., the domain where the function is non-zero) contains the support of $\probp{\sampleX}$ to perform the importance sampling of Eq.~\eqref{Eq:gradientImportanceApprox} correctly. Notice that, in order to properly optimize the parameters $\phi$, $\probp{\sampleX}$ does not need to be normalized, since this simply changes the magnitude of the gradient, but not its direction. The lack of proper normalization can be properly handled by standard neural network optimizers such as Adam \cite{Kingma2014AdamAM}.

The method described by Eq.~\eqref{Eq:gradientImportanceApprox} implies an iterative way of optimizing $\probqphi{\sampleX}$ with the following steps:
\begin{enumerate}
    \item A batch of $\sampleX$ is generated according to the current state of the neural network, $\probqphi{\sampleX}$.
    \item Using this batch, the parameters $\phi$ of the neural network are optimized via the gradient of Eq.~\eqref{Eq:gradientImportanceApprox}.
    \item This updated neural network then generates the next batch.
\end{enumerate}

\subsection{Relevance of background noise}
\label{Sec:BackgrounNoise}

As briefly discussed in previous Section \ref{Sec:Background}, in order to optimize the neural network following Eq.~\eqref{Eq:gradientImportanceApprox}, the gradient is only correctly computed if the support of $\probqphi{\sampleX}$ contains the one of $\probp{\sampleX}$. Moreover, if we want to use $\probqphi{\sampleX}$ as our proposal function to sample from $\probp{\sampleX}$ via rejection sampling, this also has to hold. 

To ensure the proper $\probp{\sampleX}$ support, we introduce the concept of a  background density function, $\probbg{\sampleX}$. In HEP, as in many other scientific areas, the density is restricted to a certain domain of $\sampleX\in\mathbb{R}^D$, e.g., the cosine has to be in $[-1,1]$, the magnitude of the momentum in an experiment has to be positive and has a maximum value of $p_\text{max}$, there are constraints in the conservation of  energy and momentum, etc... Hence $\probbg{\sampleX}$ should be a density that has a support beyond these phase-space boundaries. In what follows, a uniform distribution will be considered, with limits in each dimension according to the phase space of that coordinate. The selection of the functional form of the $\probbg{\sampleX}$ is arbitrary and it can be selected to adapt to the requirements of each project. 

The background density $\probbg{\sampleX}$ will be used for two tasks:
\begin{enumerate}[(i)]
    \item Improve initial training: At the beginning of the training, we cannot assure that the support of $\probqphi{\sampleX}$ contains the one of $\probp{\sampleX}$. Hence, instead of using $\probqphi{\sampleX}$ for the importance sampling of Eq.~\eqref{Eq:gradientImportanceApprox}, $\probbg{\sampleX}$ will be used during the warm-up phase of the training. The distribution of the weight function $w(\sampleX)=\probp{\sampleX}/\probbg{\sampleX}$ might span several orders of magnitude, but this way we ensure the full support of $\probp{\sampleX}$. This strategy gives a better approximation than the one obtained by the randomly initialized neural network $\probqphi{\sampleX}$ at the start of the training.

    \item Ensure exhaustive coverage of the phase space: The target density $\probtarget{\sampleX}$ that the neural network will learn will be constructed as a linear combination of the true target density $\probp{\sampleX}$ and the background $\probbg{\sampleX}$:
    \begin{align}\label{Eq:convexCombination}
        \probtarget{\sampleX}=(1-\alpha)\cdot\probp{\sampleX}+\alpha\cdot\probbg{\sampleX},
    \end{align}
    with $\alpha\in(0,1)$. This implementation adds a certain percentage $\alpha$ of background noise to the target density, spreading it over all the domain of the background density, allowing to properly apply the methods rejection and importance sampling with $\probqphi{\sampleX}$ as the proposal function, covering exhaustively the phase space. Experimentally we have found good compromise with $\alpha=0.05$.
\end{enumerate}

Optimizing $\probqphi{\sampleX}$ to match $\probtarget{\sampleX}$ of Eq.~\eqref{Eq:convexCombination} instead of $\probp{\sampleX}$ will make the proposal $\probqphi{\sampleX}$ slightly worse for rejection/importance sampling.  By performing the optimization to $\probp{\sampleX}$ directly in an iterative way, as explained at the end of the last Section, some regions of the phase space might disappear for future  samplings. These regions are located normally close to the boundaries of sampled volume. Having a constant background noise prevents these losses from appearing, as the neural network has to also learn to generate this noise, covering properly the required phase-space volume. We will discuss the impact of the background term on the method performance in Section \ref{Sec:MCCCQE}.

\subsection{ENIS training scheme of the proposal function}
\label{Sec:TrainingScheme}

\begin{figure*}[!thp]
\begin{center}
\includegraphics[width=.9\textwidth]{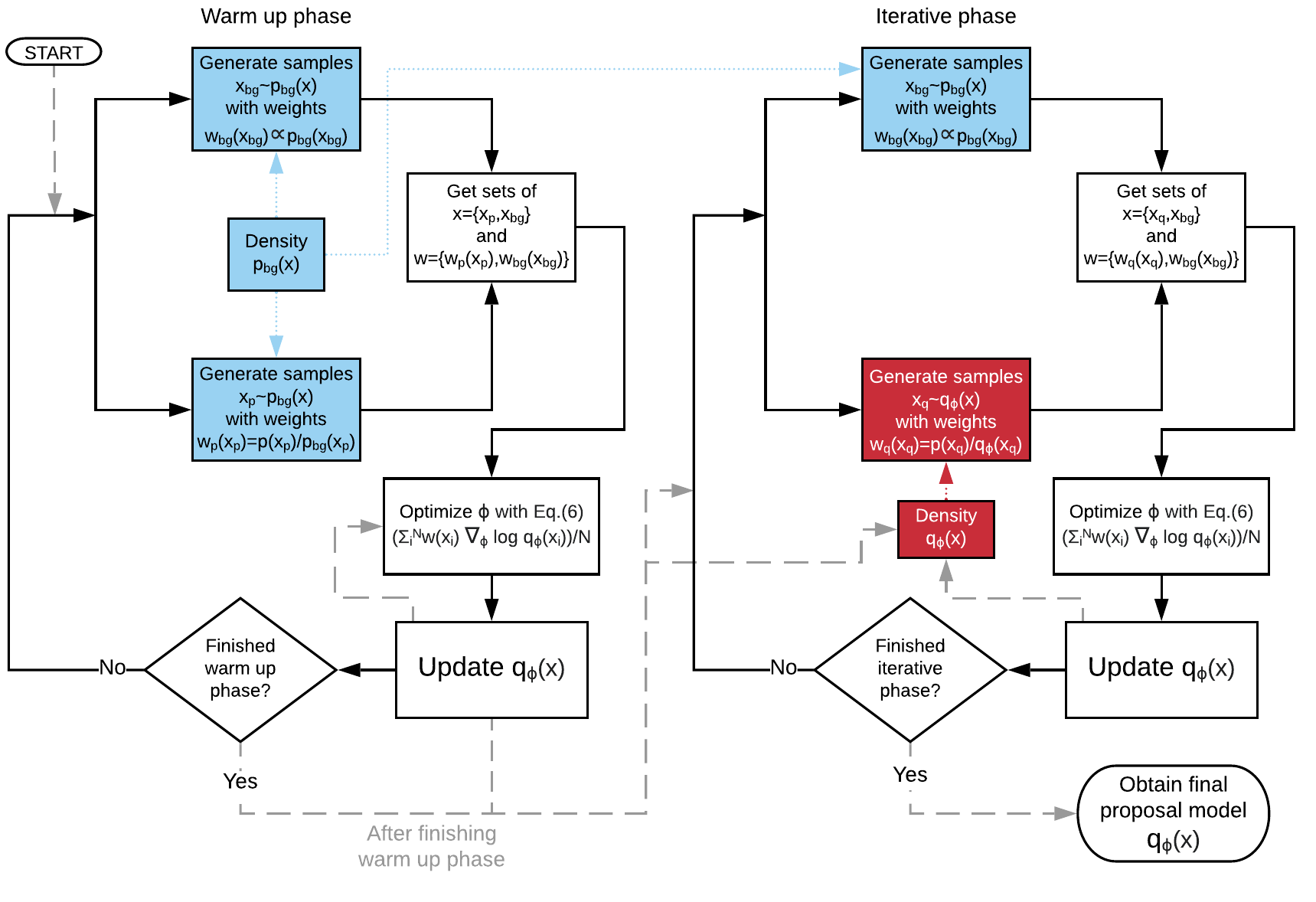}
\caption{ Exhaustive Neural Importance Sampling flow diagram. The warm-up phase is depicted on the left block and the iterative phase on the right block. See text for a detail description.}
\label{Fig:ENISScheme}
\end{center}
\end{figure*}

The training procedure to obtain  $\probqphi{\sampleX}$ from $\probp{\sampleX}$ following ENIS consists of two phases:
\begin{enumerate}
    \item Warm-up phase:
    \begin{enumerate}[(i)]
        \item Sample $\sampleX_p\sim\probbg{\sampleX}$ and compute their weights $w_p(\sampleX_p)=\probp{\sampleX_p}/\probbg{\sampleX_p}$.
        
        \item Sample background $\sampleX_\text{bg}\sim\probbg{\sampleX}$ with associated weights $w_\text{bg}(\sampleX_\text{bg})= C_{w_\text{bg}} \cdot \probbg{\sampleX_\text{bg}}$, where $C_{w_\text{bg}} = \frac{\alpha}{1-\alpha} \frac{\langle w_p (\sampleX_p) \rangle}{\langle p_\text{bg}(\sampleX_\text{bg}) \rangle}$.
        
        \item Optimize the parameters of $\probqphi{\sampleX}$ via Eq.~\eqref{Eq:gradientImportanceApprox} using $\sampleX = \{\sampleX_p,\sampleX_\text{bg}\}$ with weights $w(\sampleX) = \{w_p(\sampleX_p),w_\text{bg}(\sampleX_\text{bg})\}$.
    \end{enumerate}
        \item Iterative phase:
    \begin{enumerate}[(i)]
        \item Sample $\sampleX_q\sim\probqphi{\sampleX}$ and compute their weights $w_q(\sampleX_q)=\probp{\sampleX_q}/\probqphi{\sampleX_q}$.
        \item Sample background $\sampleX_\text{bg}\sim\probbg{\sampleX}$ with associated weights $w_\text{bg}(\sampleX_\text{bg})= C'_{w_\text{bg}}\probbg{\sampleX_\text{bg}}$, where $C'_{w_\text{bg}} = \frac{\alpha}{1-\alpha} \frac{\langle w_q (\sampleX_q)\rangle}{\langle p_\text{bg}(\sampleX_\text{bg}) \rangle}$.
        \item Optimize the parameters of $\probqphi{\sampleX}$ via Eq.~\eqref{Eq:gradientImportanceApprox} using $\sampleX = \{\sampleX_q,\sampleX_\text{bg}\}$ with weights $w(\sampleX) = \{w_q(\sampleX_q),w_\text{bg}(\sampleX_\text{bg})\}$.
    \end{enumerate}
\end{enumerate}
Fig.~\ref{Fig:ENISScheme} depicts a flow diagram of the training method for ENIS, showing on the left block the warm-up phase, while on the right block the iterative phase. The phase transition from warm-up to iterative phase is chosen heuristically. In our particular implementation of Sec.~\ref{Sec:Training} we have chosen the warm-up phase to comprise 20 \% of the total training iterations. 

Steps 1.~(ii) and 2.~(ii) allow the method to add background following Eq.~\eqref{Eq:convexCombination} to construct $\probtarget{\sampleX}$ even if $\probp{\sampleX}$ is not normalized.

\subsection{Measuring the performance of the proposal function}
\label{Sec:MeasuringPerformance}

The proposed method is not to use $\probqphi{\sampleX}$ as a direct approximation of $\probp{\sampleX}$, but as proposal function to perform either rejection (Sec.~\ref{Sec:RejectionSampling}) or importance sampling \cite{10.2307/166789}. This allows for the methods to correct any deviation in the neural network modeling of the exact density while utilizing its proximity to such density.

We use the learned probability density function $\probqphi{\sampleX}$ to generate samples via rejection sampling (see Sec.~\ref{Sec:RejectionSampling}), which, in HEP, is of high interest and costly via standard procedures. The parameter $k$ of the rejection algorithm has to be estimated empirically. Consider $n$ samples $\{\sampleX_i\}_{i=1}^n$ generated with the proposal function $\sampleX\sim\probq{\sampleX}$, with weights $w(\sampleX_i)=\probp{\sampleX}/\probq{\sampleX}$, satisfying:
\begin{itemize}
    \item The average of the weights is
    \begin{align}
        \langle w\rangle = \frac{1}{N} \sum_{i=1}^N w(\sampleX_i)\approx \int \probq{\sampleX} w(\sampleX) d\sampleX = C,
    \end{align} 
    where $C$ is the normalization of the density $\probp{\sampleX}$, i.e., its volume.
    \item $\kmax$, the smallest constant $k>0$ such that the inequality of Eq.~\eqref{Eq:rejection} holds, is equal to $(\max w(\sampleX_i))^{-1}$.
\end{itemize}

In real conditions, the parameter $k$ can be relaxed. Instead of choosing the maximum value among the empirically computed weight distribution, it can be taken as the inverse of the $Q$-quantile of these weights, $\wQ{}$, denoted by $\kQ{}$:
\begin{align}\label{Eq:k-quantile}
    \kQ{} = (Q\text{-quantile}(w))^{-1} = \wQ{}^{-1}.
\end{align}
This is equivalent of clipping the weights' maximum value to the $Q$-quantile of $w$, capping the desired density function $\probp{\sampleX}$ we are generating using these weights for the rejection. The new weights $w'(\sampleX)$ are simply:
\begin{equation}
    w'(\sampleX_i)=\left\{
    \begin{array}{ll}
        w(\sampleX_i) &\text{if } w(\sampleX_i)\leq \wQ{} \\
        \wQ{} &\text{if } w(\sampleX_i)> \wQ{} 
    \end{array}
    \right..
\end{equation}
The ratio of volume with respect to the original density $\probp{\sampleX}$ we are maintaining by clipping the weights this way defines the coverage we have of the rejection sampling, and is equal to
\begin{align}\label{Eq:Coverage}
    \text{Coverage} = \frac{\sum_{i=1}^N w'(\sampleX_i)}{\sum_{i=1}^N w(\sampleX_i)}.
\end{align}
This allows us to quantify the error we are committing when choosing a quantile over the maximum of weights when defining a constant $k$ for rejection sampling.

The idea behind relaxing this constant $k$ is that we will approximate wrongly only a small region of $\probp{\sampleX}$ with $\probq{\sampleX}$. In that small region, the ratio $\probp{\sampleX}/\probq{\sampleX}$ is large compared to the rest of the domain but still it is occupying a small volume of the density $\probp{\sampleX}$. This region can be ignored by relaxing $k$, making the overall ratio of $\probp{\sampleX}/(k\cdot \probq{\sampleX})$ closer to 1 and improving drastically the rejection sampling at the cost of this small discrepancy which we are committing, quantified in Eq.~\eqref{Eq:Coverage}.

As an additional qualitative measurement of the goodness of different proposals under different constants $k$, the effective sample size (ESS) will be used \cite{Liu1996}, which corresponds approximately to the number of independent samples drawn. The ESS for $n$ samples of weights $\{w(\sampleX_i)\}_{i=1}^n$ is defined as:
\begin{align}\label{Eq:ness}
    N_\text{ESS} = \left(\sum_{i=1}^N w(\sampleX_i) \right)^2/\sum_{i=1}^N w(\sampleX_i)^2.
\end{align}
This is a rule of thumb to obtain the number of independent samples. The closer $N_\text{ESS}$ is to the number of samples $n$, the more uncorrelated the weighted samples are. If large weights appear, then the independence of the samples will be diminished, as a same sample gets represented many times. We define $N_\text{ESS}/N$ as a rough estimate for the ratio of independence of the samples.

\section{Monte Carlo generation of the CCQE antineutrino cross-section}
\label{Sec:MCCCQE}

We will now proceed to apply ENIS to the CCQE antineutrino cross-section density. In Sec.~\ref{Sec:Training}, we discuss how the training for the NSF was performed, describing the background we added to cover the phase space. We show qualitatively the obtained densities and compare them to the target one. After obtaining a suitable proposal, we discuss in depth the performance of the obtained result in Sec.~\ref{Sec:PerformanceAndDiscussion}, comparing the ENIS proposal to a generic uniform one, demonstrating its potential while justifying the relaxation on the constant $k$ for the rejection sampling.

\subsection{Training} 
\label{Sec:Training}

To find the proposal function $\probqphi{\sampleX}$ via NSF for the CCQE antineutrino interaction cross-section density, described in Sec.~\ref{Sec:CCQEXsect}, we followed the training scheme from Sec.~\ref{Sec:TrainingScheme}.

The background chosen is a uniform distribution, covering a range of $[0,10]$ for the incoming neutrino energy $E_\nu$ (in GeV), $[0,10]$ for the outgoing muon momentum $p_\mu$ (in GeV/c), $[0,\pi]$ for the angle of the outgoing muon $\theta_\mu$ (in rad), and $[0,0.225]$ for the target nucleon Fermi momentum $p_\mathrm{nucleon}$ (in GeV/c). These bounds were expanded by covering a slightly more extended domain, of an additional $2$~\% at the beginning and end of each dimension, to assure that the physical boundaries are completely covered. This expanded background was added with a $\alpha=0.05$ contribution to the cross-section density in Eq.~\eqref{Eq:convexCombination}, as well as used during training for the warm-up phase.

As for the hyperparameters of the NSF, we have chosen the \textsc{Adam} optimizer \cite{Kingma2014AdamAM} with learning rate $0.0005$, batch size $5\,000$, training steps $400\,000$, $5$ flow steps, $2$ transform blocks, $32$ hidden features and $8$ bins. This gives a total dimension of $37\,220$ for the parameters $\phi$ of $\probqphi{\sampleX}$. This configuration for the NSF was chosen experimentally to have a relatively low number of parameters (one can have easily six million parameters instead of the $\approx 37\,000$ we have) since a lower number speeds up the generation and evaluation of samples $\sampleX\sim\probqphi{\sampleX}$. Additionally, the learning rate was decreased during the training using a cosine scheduler to ensure stabilization at the end of the training procedure.

\begin{figure}[!thp]
\begin{center}
\includegraphics[width=\columnwidth]{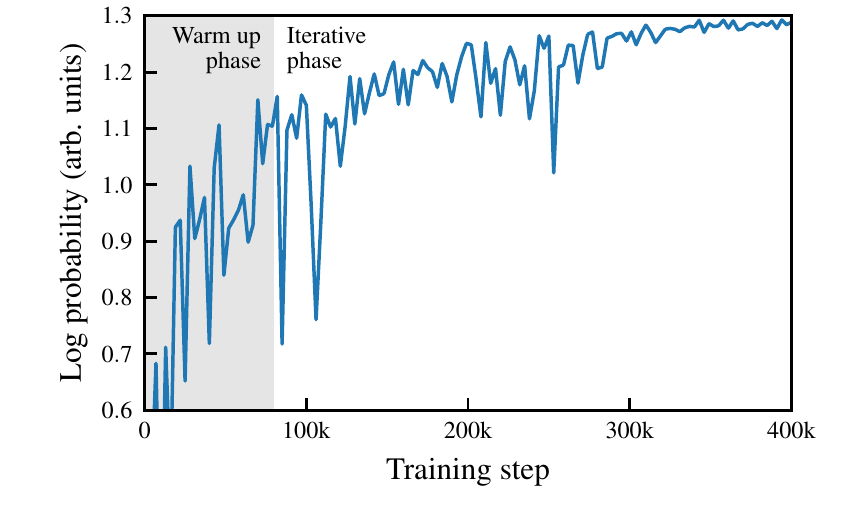}
\caption{Neural Spline Flow training log probability for estimating the modified CCQE cros-section with background noise, following Eq.~\eqref{Eq:objective}. In grey, the warm-up phase is performed, using $\probbg{\sampleX}$ to generate the weighted samples, while in the white area the current state of the NSF $\probqphi{\sampleX}$ is used. The log probability stabilizes during the training to converge to a certain value which depends on the expressiveness of the network and the normalization of the target density.}
\label{Fig:trainLog}
\end{center}
\end{figure}

The training consists in maximizing the log-likelihood of Eq.~\eqref{Eq:objective} by computing its gradient via Eq.~\eqref{Eq:gradientImportanceApprox}, and is shown over the $400\,000$ iterations in Fig.~\ref{Fig:trainLog}. In the grey area, the training is performed with samples of the background distribution $\probbg{\sampleX}$, while in the white area the samples of the training samples are generated by the current proposal distribution $\probqphi{\sampleX}$. Notice that since the samples are generated in real-time during the training, there is no need to worry about possible overfitting of the parameters of the neural network, which is a common issue in many machine-learning applications. The values of Fig.~\ref{Fig:trainLog} are computed every one thousand steps, for a batch of $200\,000$ samples $\sampleX\sim\probqphi{\sampleX}$. The log probability can be seen to converge at the end of the training, which is mainly due to the cosine scheduler, but also due to the saturation over the family of parametrized densities $\probqphi{\sampleX}$.

\begin{figure}[!thp]
\begin{center}
\includegraphics[width=\columnwidth]{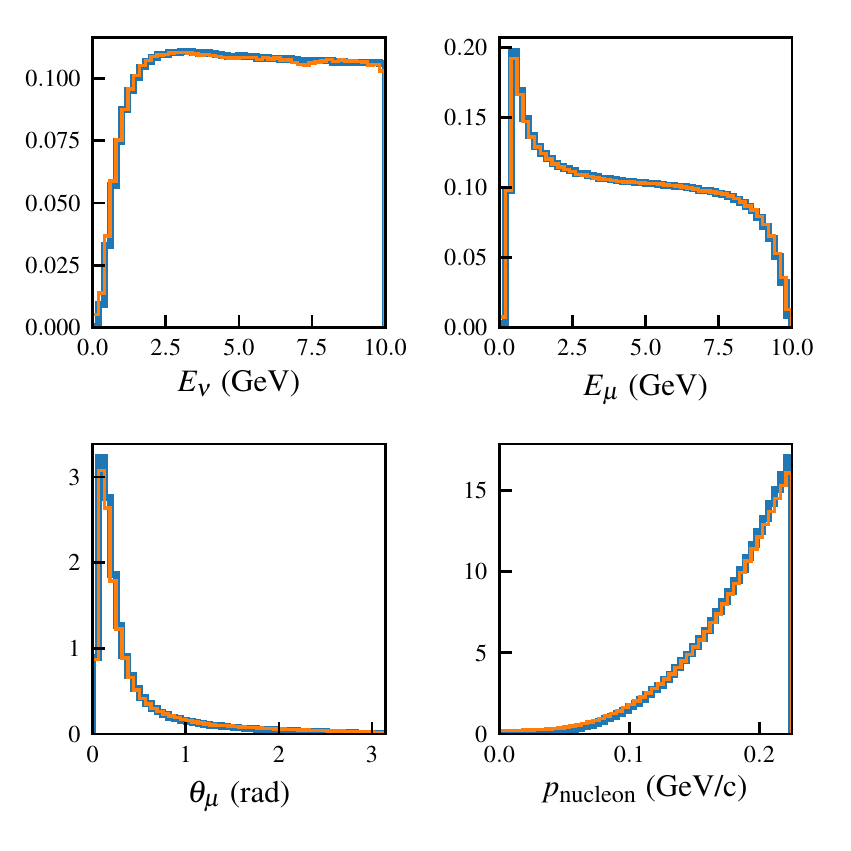}
\caption{$\probp{\sampleX}$ (blue) vs $\probqphi{\sampleX}$ (orange) 1-dimensional normalized histograms of the marginalized CCQE cross-section density for each of the variables. The plots show light discrepancy in each variable, but an overall agreement between the NSF proposal $\probqphi{\sampleX}$ and the CCQE cross-section density $\probp{\sampleX}$. Notice how the distribution of $\probqphi{\sampleX}$ are taken before performing rejection sampling on it.}
\label{Fig:1Ddenisty}
\end{center}
\end{figure}

To have a visual representation, Fig.~\ref{Fig:1Ddenisty} shows the marginalized 1-dimensional densities of the four cross-section variables of the target density $\probp{\sampleX}$ (blue) vs the NSF proposal $\probqphi{\sampleX}$ (orange). The plots show a small discrepancy in each variable, but an overall agreement between the two densities. Aside from a mismodeling on the side of $\probqphi{\sampleX}$ in certain regions, the differences can also come from the fact that the NSF is learning a modified target density (Eq.~\eqref{Eq:convexCombination}).

\begin{figure*}[!thp]
\begin{center}
\begin{minipage}[c]{\columnwidth}
$\probp{\sampleX}$
\includegraphics[width=\columnwidth]{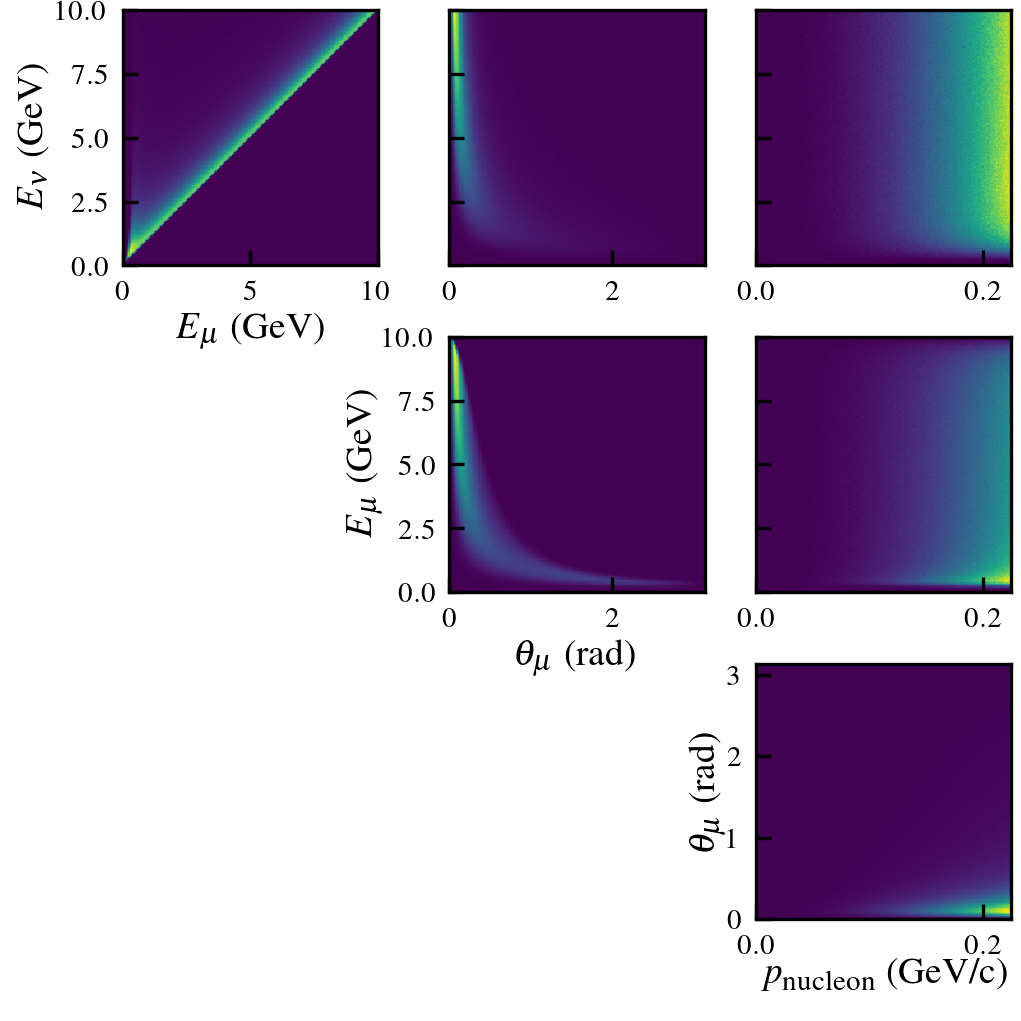}
\end{minipage}
\begin{minipage}[c]{\columnwidth}
$\probqphi{\sampleX}$
\includegraphics[width=\columnwidth]{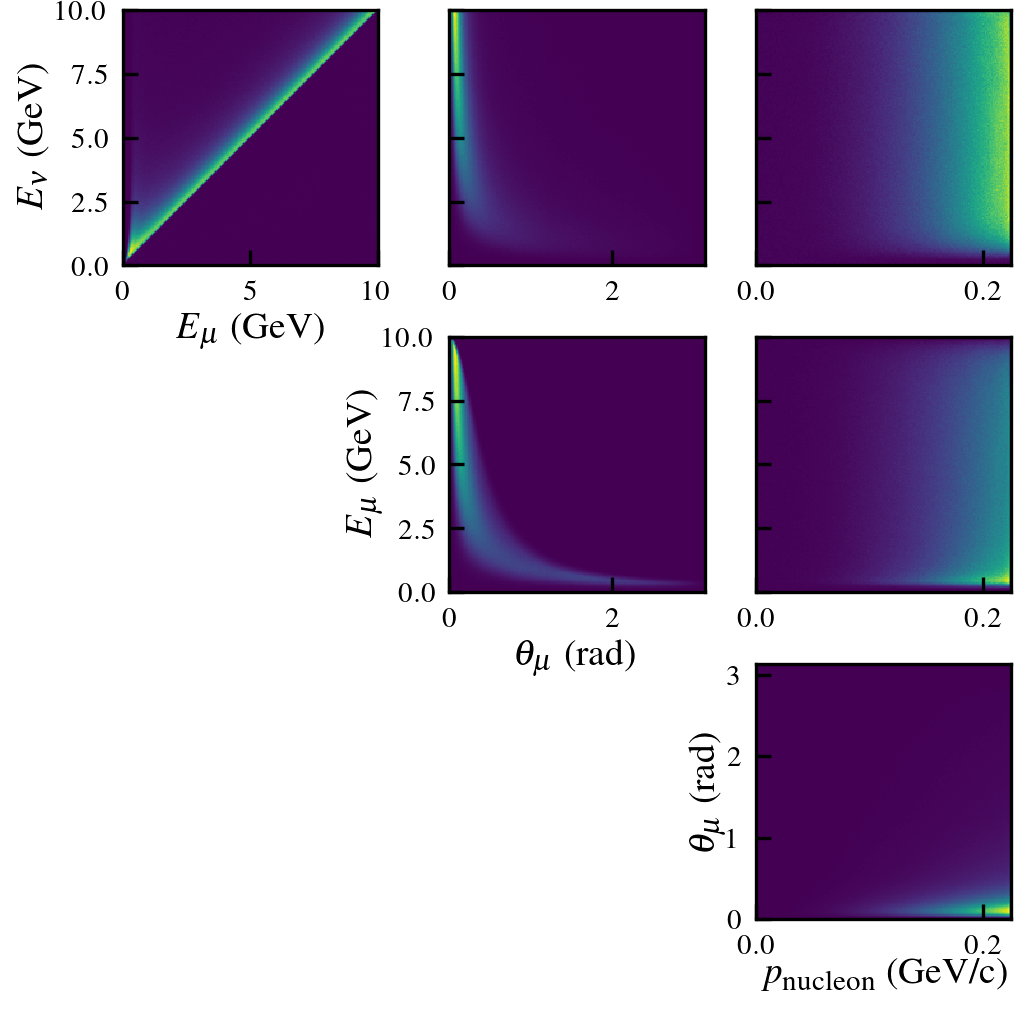}
\end{minipage}
\caption{2D histograms comparison of cross-section density for the real cross-section $\probp{\sampleX}$ (left) and the proposal density $\probqphi{\sampleX}$(right). Visually, an overall agreement can be seen.}
\label{Fig:2Ddenisty}
\end{center}
\end{figure*}

To asses qualitatively that the correlation between the variables are also captured by $\probqphi{\sampleX}$, Fig.~\ref{Fig:2Ddenisty} shows 2D-histograms for both the real density $\probp{\sampleX}$ (left) and the the proposal density $\probqphi{\sampleX}$ (right).  Visually, an overall agreement can be seen. There is a slight discrepancy for high energy $p_\mathrm{nucleon}$ values, where the attenuation indicates that for the NSF proposal function it is more spread due to the background noise $\probbg{\sampleX}$ it is also learning (Eq.~\eqref{Eq:convexCombination}).

In what follows the performance of the NSF proposal will be discussed in more quantitative ways, and compared it to a uniform proposal.

\subsection{Performance and discussion} 
\label{Sec:PerformanceAndDiscussion}

In this section we will focus on analyzing the performance of the proposal density obtained by the NSF while also comparing it to a uniform proposal density, $\probunif{\sampleX}$, which in our case will be the same as $\probbg{\sampleX}$, defined in Sec.~\ref{Sec:Training}.

\begin{figure}[!thp]
\begin{center}
\includegraphics[width=\columnwidth]{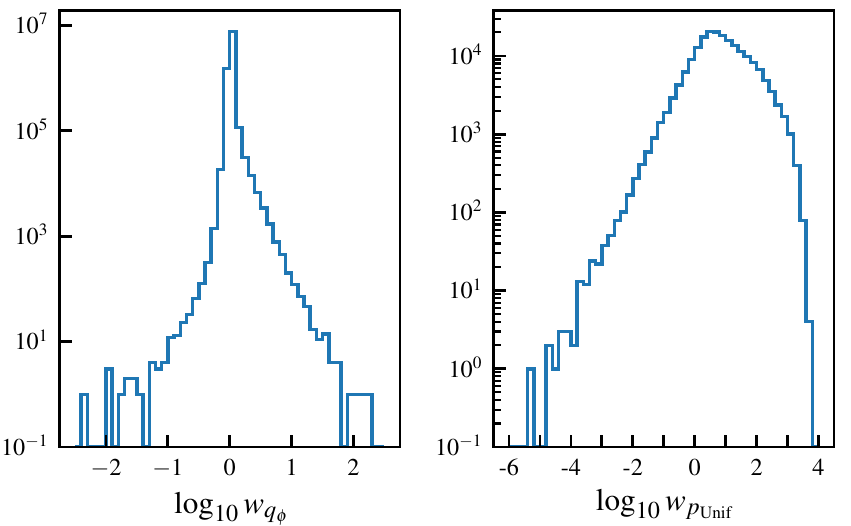}
\caption{Logarithmic weight distribution for ten million samples from the NSF proposal $\probqphi{\sampleX}$ (left) vs the same number of samples from the uniform proposal $\probunif{\sampleX}$ (right). Notice that we are only computing the logarithm for weights $>$ 0. For the NSF, all weights are concentrated around $\log_{10} w=0$ with a small dispersion around it, while for the uniform distribution the spectrum of weights goes over 9 orders of magnitude.}
\label{Fig:logw}
\end{center}
\end{figure}

We start by generating ten million samples from each proposal density and compute their associated weights. The proportion of samples with weight equal to zero is $5.56$~\% for the NSF proposal, compared to the $98.03$~\% for the uniform one. To understand the distribution of such weights, Fig.~\ref{Fig:logw} shows the logarithmic scale of them (for the weights $>$ 0), assuming the average of the weights is equal to 1. For the NSF $\probqphi{\sampleX}$ (left), all weights are concentrated around $\log_{10} w=0$ with a small dispersion around it. Notice that there are only three weights in ten million slightly over one hundred. This shape justifies using not the maximum value of $w$ to perform rejection sampling, but some quantile of it, as we will discuss below. Contrary, for the uniform distribution $\probunif{\sampleX}$ (right) we can see that the spectrum of weights goes over nine orders of magnitude. The mean for $\log_{10} w_{q_\phi}$ is $0.023\pm0.040$, while for $\log_{10} w_{p_\mathrm{Unif}}$ we obtain an average of $0.85\pm0.88$, indicating a huge fluctuation in the magnitude of the weights.

The results of the performance test for rejection sampling are summarized in Tab.~\ref{Tab:Performance}, where we compare various quantities for the NSF $\probqphi{\sampleX}$ and uniform $\probunif{\sampleX}$ proposal functions. For this, different quantiles for the constant $k$ for the rejection method are used, following Eq.~\eqref{Eq:k-quantile}, relaxing its restriction as discussed in Sec.~\ref{Sec:MeasuringPerformance}. The quantiles for $k$ were chosen using the ten million weights computed for the previous discussion of the weight magnitudes, as well as the probability of acceptance, the coverage, and the effective sample size. We considered a case of sampling one million accepted samples via rejection sampling, where samples from the proposal were generated and checked for acceptance/rejection in parallel, in batches of $300\,000$ samples. The purpose of the parallelization is to exploit the computational capacities of a GPU. We denoted each of these batches of generating and checking a cycle of the rejection sampling. The values in Tab.~\ref{Tab:Performance}, for each quantile value and a proposal function, are the following:
\begin{itemize}
    \item $p_\mathrm{accept}$: probability of accepting a single event, given by the average of $\probp{\sampleX}/(k\cdot \probqphi{\sampleX})$. If $\probp{\sampleX}/(k\cdot \probqphi{\sampleX})>1$, it is taken as 1 for the computation.
    \item Cycles: number of rejection sampling cycles of size $300\,000$ samples needed to obtain one million accepted samples: 
    \begin{align}\label{Eq:cycles}
        \text{Cycles}=\Bigl\lceil\frac{10^6}{p_\mathrm{accept}\cdot3\times10^5}\Bigr\rceil
    \end{align}
    \item Time: seconds it takes to compute these cycles and obtain one million accepted samples: $t_\mathrm{cycle}\cdot \text{Cycles}$.
    \item Coverage: volume of the original density covering when taking $k$ with a certain quantile (Eq.~\eqref{Eq:k-quantile}), following Eq.~\eqref{Eq:Coverage}.
    \item $N_\mathrm{ESS}/N$: the ratio of effective sample size over the total number of samples, quantifying an estimate of the ratio of independence of the events. This was computed for a sample size of 10 million.
\end{itemize}

\begin{table}[!htp]
\caption{Performance values for different quantile choices of $k$ for rejection sampling, as discussed in Sec.~\ref{Sec:MeasuringPerformance}, comparing both NSF $\probqphi{\sampleX}$ and uniform $\probunif{\sampleX}$ proposal functions. For this exercise, one million samples were generated, performing rejection sampling in batches of $300\,000$ tries.  The quantities are the probability of accepting a single sample $p_\mathrm{accept}$, the number of rejection cycles (batches of $300\,000$) used to obtain one million accepted samples, the time it took in seconds to generate these accepted samples, the coverage of the target density for that particular quantile (Eq.~\eqref{Eq:Coverage}) and the ratio of effective sample size $N_\mathrm{ESS}$ (Eq.~\eqref{Eq:ness}) over the total number of samples $N$.}
\label{Tab:Performance}
\begin{tabular}{l|l|l|l|l|l|l}
\toprule
Quantile & Prop. & $p_\mathrm{accept}$  &  Cycles  &  Time (s) & Coverage &  $N_\mathrm{ESS}/N$\\
\hline
1.00000 &  NSF &  0.0051 &   649 & 201.822 & 1.0000 &   0.9140\\
& Unif. & 0.0002 & 16199 & 47.886 & 1.0000 &   0.0016\\
\hline
0.99999 & NSF &   0.0633 &   53 & 16.482 &   0.9999 &  0.9242 \\
& Unif. & 0.0004 & 8056 & 23.814 & 0.9947 &  0.0017\\
\hline
0.99990 &  NSF & 0.1623 &  21 &  6.530 &  0.9996 &  0.9284 \\
& Unif. & 0.0008 &  4240 &   12.534 &  0.9480 &  0.0020\\
\hline
0.99900 & NSF & 0.3590 &  10  & 3.110 & 0.9984 & 0.9338\\
& Unif. & 0.0027 & 1217 &  3.598 & 0.6185 & 0.0045\\
\hline
 0.99000 &  NSF & 0.7187 & 5  & 1.555 & 0.9939 & 0.9400\\
 & Unif. & 0.0137 & 244 & 0.721 & 0.0818 & 0.0154\\
 \hline
 0.98500 & NSF & 0.7730 & 5 & 1.555 & 0.9927 &  0.9405\\
 & Unif. & 0.0171 & 195 &  0.576 & 0.0325 & 0.0179\\
 \hline
 0.98100 & NSF & 0.7968 & 5 & 1.555 & 0.9920 & 0.9408 \\
 & Unif. & 0.0193 & 173 & 0.511 & 0.0039 &  0.0195
\end{tabular}
\end{table}

The probability of acceptance, $p_\mathrm{accept}$, for the NSF is at least one order of magnitude higher than the one obtained from uniform sampling. Additionally, NSF grows rapidly towards $\sim70$~\% acceptance while also covering $>99$~\% of the original density volume, as shown in the Coverage column. This is not the case for the uniform distribution, which, while being only one order of magnitude behind NSF with regards to acceptance, is missing a large volume of coverage of the original density. 

The number of rejection sampling cycles needed to achieve the desired number of accepted samples is inversely proportional to $p_\mathrm{accept}$, as shown in Eq.~\eqref{Eq:cycles}. In a cycle, the algorithm has to sample from the proposal and evaluate both the proposal and $\probp{\sampleX}$. For the NSF, the cycles get stalled when reaching a high percentage of acceptance, since the number of cycles has to be a whole number, which is equivalent to the whole number +1. Notice how the number of cycles for the NSF is two orders of magnitude smaller compared to the uniform one, however, the coverage of the uniform drops drastically when decreasing the quantile, and hence the quality of the samples. We will discuss more in-depth in Appendix~\ref{Sec:appendixComplexity}.

When looking at the time it takes to obtain one million accepted samples, it is directly proportional to the cycles for each proposal. The main difference is that a cycle for the uniform proposal takes a fraction of the time of a cycle for the NSF. This is because sampling and evaluating for the NSF is heavy computationally compared to doing this task for a uniform distribution. As mentioned before, see Appendix~\ref{Sec:appendixComplexity} for a more in-depth discussion. 

The coverage is the main quantity of measurement of the quality of the produced samples since it measures the volume conserved of the original distribution when performing rejection sampling with certain quantiles. For all the chosen quantiles, the NSF drops a volume $<1$~\%, while for the uniform distribution the loss is of $>5$~\% for quantile $0.9999$, $>38$~\% for $0.999$, and $>91$~\% for $0.99$, which is unacceptable when trying to produce samples from the original distribution. For the NSF this level of performance when taking the above quantiles is expected, as in Fig.~\ref{Fig:logw} we have seen that the upper tail of weights with large magnitudes is a small percentage of the whole distribution. However, for the uniform distribution, the loss of coverage is caused by two facts: (i)  $98.03$~\% of the weights are zero, hence placing the whole distribution on a $1.97$~\% of the weights. (ii) These weights, as seen in Fig.~\ref{Fig:logw}, span over many orders of magnitude, making a cut on the quantile of their distribution more noticeable, as will be discussed below.

\begin{figure*}[!tp]
\begin{center}
\begin{minipage}[c]{\columnwidth}
Marginalized coverage of $\probqphi{\sampleX}$
\includegraphics[width=\columnwidth]{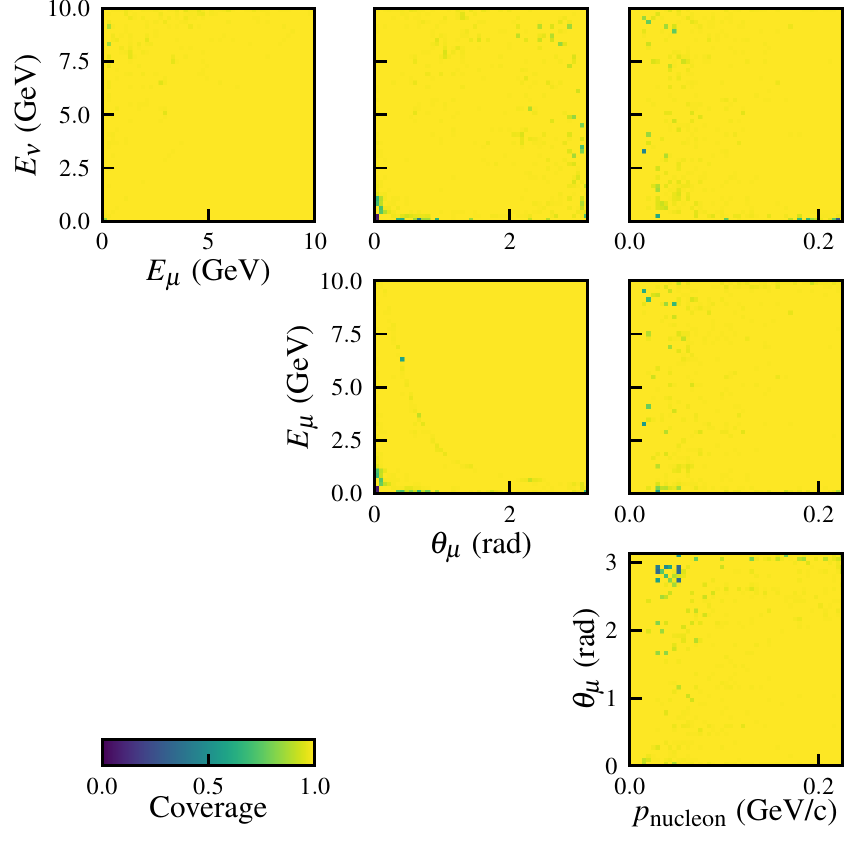}
\end{minipage}
\begin{minipage}[c]{\columnwidth}
Marginalized coverage of $\probunif{\sampleX}$
\includegraphics[width=\columnwidth]{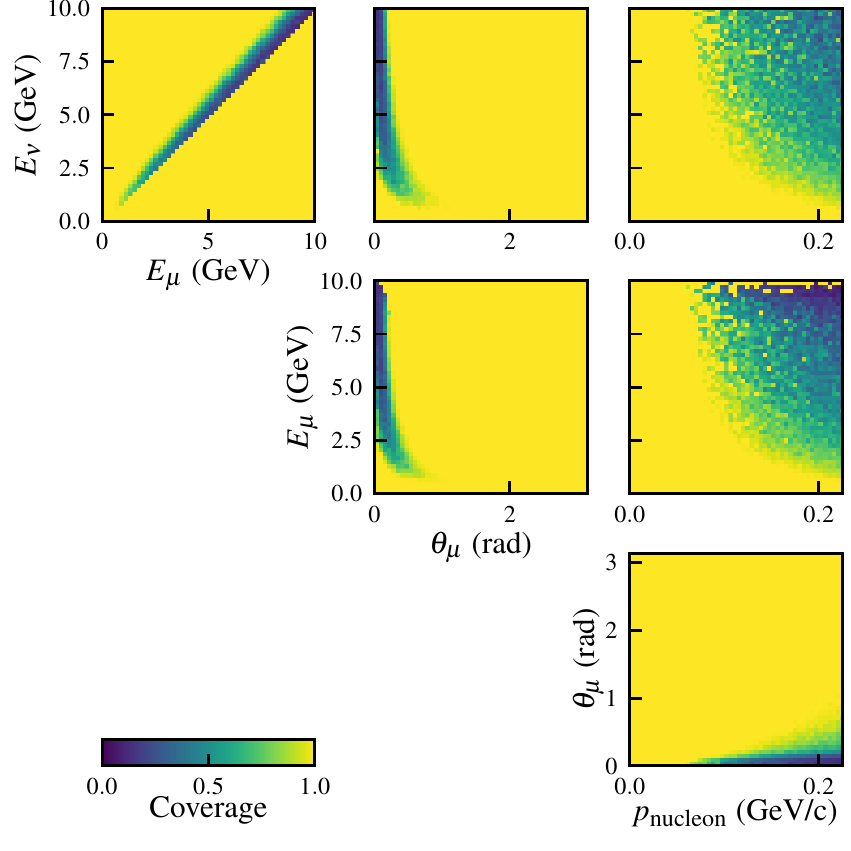}
\end{minipage}
\caption{2D histogram representation of the marginalized coverage bin-to-bin for NSF proposal (left) and for the uniform proposal (right), for $k$-quantile$=0.999$. The coverage of the NSF presents a negligible discrepancies in small areas, justifying the use of a quantile for $k$ to improve acceptance and time, as shown in Tab.~\ref{Tab:Performance}. For the uniform proposal, the coverage presents an important size of the total area with significant low coverage, which is unacceptable when trying to perform rejection sampling from it.}
\label{Fig:2DCoverage}
\end{center}
\end{figure*}

To visualize the coverage and the regions missing by choosing a quantile $\kQ{0.999}$ and different proposal functions, Fig.~\ref{Fig:2DCoverage} shows 2-dimensional histogram representation of the marginalized coverage bin-to-bin, taking the variables in pairs, where each bin quantifies the coverage of that bin (i.e., the sum of weights in that bin after choosing a certain quantile over the sum of weights of those weights without clipping). On the left, the coverage for the NSF is presented and shows that only few regions of the phase space have values smaller than 1, and even in those regions the coverage has no noticeable discrepancies. On the right, the coverage of the uniform proposal is shown for the same quantile $0.999$, marking clear regions where the coverage drops drastically to values close to zero. 

When comparing both coverage regions in Fig.~\ref{Fig:2DCoverage}, a clear pattern be seen for the uniform one, while it looks quite random for the NSF. This is because the coverage is related to the ratio $\probp{\sampleX}/\probq{\sampleX}$, with $\probq{\sampleX}$ the corresponding proposal density. For the NSF, $\probqphi{\sampleX}$ has a shape very closely related to $\probp{\sampleX}$, as shown in Fig.~\ref{Fig:1Ddenisty} and Fig.~\ref{Fig:2Ddenisty}, so the coverage would correspond to regions where the discrepancy is large, which has a noisy behavior. Contrary, for the uniform proposal, $\probunif{\sampleX}$, this ratio is proportional to $\probp{\sampleX}$, hence, by clipping, we are doing so according to that particular shape, making the coverage less chaotic and more structured. This translates into making highly probable areas equally likely than others with less probability, affecting this exact group of regions as we will now analyze.

\begin{figure}[!thp]
\begin{center}
\includegraphics[width=\columnwidth]{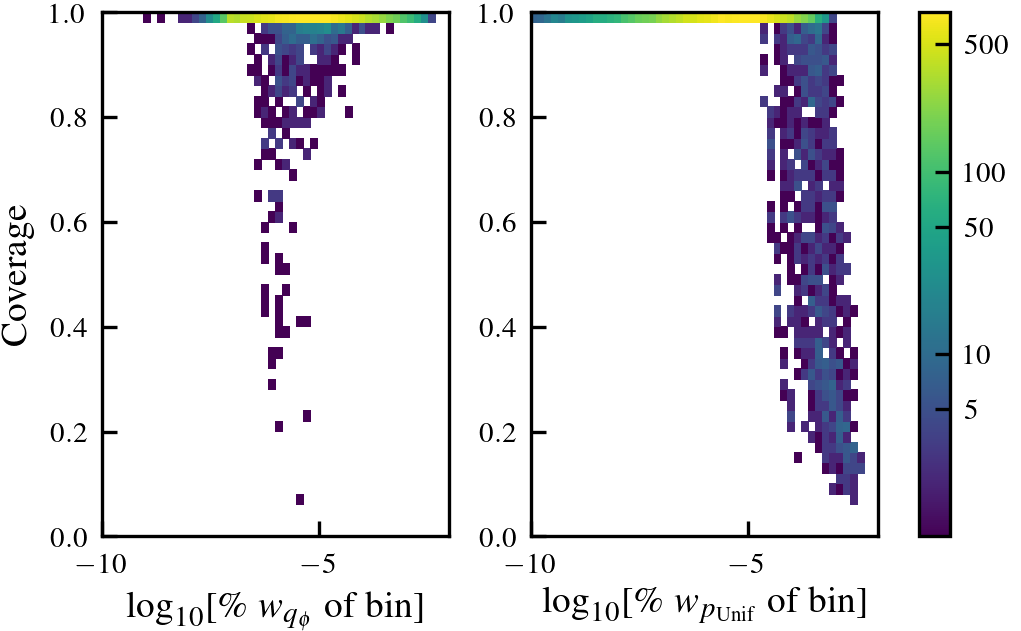}
\caption{Representation of the number of bins according to their $\% w_q$ of bin vs their coverage for four-dimensional bins in $\probp{\sampleX}$, taking $\kQ{0.999}$. For the NSF (left), the coverage is close to one in most of the bins, and when it drops it is for low $\% w_q$. Contrary, for the uniform proposal (right), the coverage drops for high $\% w_q$, making the overall coverage way smaller than the one for NSF, as shown in Tab.~\ref{Tab:Performance}.}
\label{Fig:binnedCoverage}
\end{center}
\end{figure}

Fig.~\ref{Fig:2DCoverage} gives us an overall picture of where the densities are wrongly estimated by choosing certain quantile, but it does not quantify or indicate the amount of error, that is, it is not telling us whether the coverage is poor in areas of small or high density. To answer this question, a multidimensional histogram over all four dimensions was performed, with 20 bins in each dimension. Then, for each bin,  we compute the percentage of weight for a proposal $q$,
\begin{align}
    \text{\% } w_q \text{ of bin } = \sum_{\sampleX\in\text{bin}} w_q(\sampleX) / \sum_{\sampleX} w_q(\sampleX),
\end{align}
which is equivalent to the percentage of density $\probp{\sampleX}$ in that bin, and the coverage for the quantile $\kQ{0.999}$. Fig.~\ref{Fig:binnedCoverage} shows a histogram of the number of bins according to their \% $w_q$ of bin vs their coverage. Notice how \% $w_q$ is presented on a logarithmic scale. For the NSF (Fig.~\ref{Fig:binnedCoverage} left), the regions of coverage visibly smaller than one are two to four orders of magnitude smaller in \%~$w_q$ than the denser high \%~$w_q$ region on the top right. This means that the areas being misrepresented by taking the quantile $\kQ{0.999}$ are relatively small. Also, most of the area with coverage $<1$ are close to full coverage. Contrary, the uniform proposal (Fig.~\ref{Fig:binnedCoverage} right) shows the coverage dropping for high values of \%~$w_q$, indicating that important regions of the original density are being trimmed down by choosing $\kQ{0.999}$. This observation is in agreement with the total coverage we are seeing in Tab.~\ref{Tab:Performance}. 

Concerning the ratio of ESS, we can see that for the NSF it is larger than $90$~\%, giving highly uncorrelated events. For the uniform proposal however, the ESS drops to the range of $0.16-1.95$~\%, even for lower quantiles. The differences can be understood from Eq.~\eqref{Eq:ness} and by looking at the weight distribution for each proposal in Fig.~\ref{Fig:logw}. A large percentage of NSF weights have the same order of magnitude while for the uniform we go over a spectrum of 8 orders of magnitudes. Additionally this ratio of ESS has to be considered for the area of the original distribution given by the coverage, where the uniform distribution poorly reproduces important regions of phase space by choosing smaller quantiles.

To summarize the analysis performed on the results of Tab.~\ref{Tab:Performance}, in the case of the NSF, by lowering the quantile, the probability of acceptance grows until reaching almost $80$~\%, reducing the time to a $0.7$~\% of the original one while maintaining coverage of over $0.99$~\%. These scores allow us to justify using a smaller quantile for rejection sampling to improve significantly the performance in time, while also quantifying the misrepresentation we are doing by lowering the constant $k$. On the contrary, for the uniform proposal, the analysis showed weakness when trying to utilize smaller quantiles, lowering the coverage by over $38$~\% when using a quantile of only $0.999$. Additionally, looking at the $N_\mathrm{ESS}$ and  Fig.~\ref{Fig:logw}, we can see that most of the distribution is concentrated in a relatively small number of samples.

\section{Conclusion}

In this letter we have presented Exhaustive Neural Importance Sampling (ENIS), a framework to find accurate proposal density functions in the form of normalizing flows. This proposal density is subsequently used to perform rejection sampling on a target density to obtain MC data sets or compute expected values. We argue that ENIS solves the main issues associated with rejection sampling algorithms as described in the introduction: (i) The training to find a good proposal is done automatically from the target density, with little configuration needed from the human point of view. (ii) Compared to generic proposal functions such as the uniform one, the normalizing flow adapts its density over the target one, getting rid of the inefficiencies which are usually on the downside of the method. (iii) The proposal function is generated based on a set of trivial normally distributed random numbers transformed through the flow, without any rejection method applied. 

The performance of the method has been demonstrated and analyzed in a real case scenario, the CCQE cross-section of the antineutrino interaction with a nucleus, where the density is four-dimensional. We have shown that, for the normalizing flow proposal, we can relax the condition in the constant $k$, used to construct the comparison function, boosting greatly the efficiency (up to $\approx 80$~\% of acceptance rate) while committing a very small error on the target density (less than a $1$~\%), bringing orders of magnitude of speed up in computing time compared to the same error committed by a uniform proposal. Additionally, we investigated the coverage of the generation method as a function of the constant $k$. We showed that the areas of the phase space where the error is committed are less relevant to the final result compared to the error in the uniform case.

The method can be used to generate fast MC samples in cases where the precision is less relevant versus the algorithm speed. High Energy Physics presents some of these examples such as extensive statistical studies based on ``Asimov data sets"\footnote{The concept of the Asimov data is being a representative date set, inspired by the short story Franchise by Isaac Asimov, in which the entire electorate is replaced by selecting the most representative voter.}, fast detector simulations, or simply in fast studies for detector developments and designs. In those cases, the learned proposal function might be sufficiently precise and easy to generate with a set of simple normal random generators transformed through the flow.

Regarding its usage, ENIS brings the possibility of applying the same normalizing flow for rejection sampling of similar densities, e.g., densities coming from a model where the parameters are changed slightly, altering the overall density smoothly. The weight distribution of the ratio between target and proposal will be altered, but no additional training of the neural network would be needed regarding the theoretical model remains similar to the original one. This is a significant advantage compared to methods like MCMC, where one would have to rerun the complete algorithm to obtain samples from each of the different densities.

As a last remark, we have demonstrated the potential of ENIS for the four-dimensional CCQE interaction density. We believe this will only be more noticeable when applying it to higher dimensions, as the original paper of NSF \cite{Durkan2019NeuralSF} shows a remarkable performance on spaces of dimensions up to sixty-three. The advantages will be also more obvious as the underlying cross-section model becomes more complex and computationally involved. 

\begin{acknowledgments}
The authors have received funding from the Spanish Ministerio de Econom\'{i}a y Competitividad (SEIDI-MINECO) under Grants No.~FPA2016-77347-C2-2-P and SEV-2016-0588, from the Swiss National Foundation Grant No. 200021\_85012 and acknowledge the support of the Industrial Doctorates Plan of the Secretariat of Universities and Research of the Department of Business and Knowledge of the Generalitat of Catalonia. The authors also are indebted to the Servei d'Estad\'istica Aplicada from the Universitat Aut\`onoma de Barcelona for their continuous support. The authors would also like to thank C. Durkan et al. for the code provided in their paper \cite{Durkan2019CubicSplineF} which served as a basis to build ENIS, and Tobias Golling for the valuable discussion he brought to the table.
\end{acknowledgments}

\bibliography{references}

\begin{thebibliography}{30}%
\makeatletter
\providecommand \@ifxundefined [1]{%
 \@ifx{#1\undefined}
}%
\providecommand \@ifnum [1]{%
 \ifnum #1\expandafter \@firstoftwo
 \else \expandafter \@secondoftwo
 \fi
}%
\providecommand \@ifx [1]{%
 \ifx #1\expandafter \@firstoftwo
 \else \expandafter \@secondoftwo
 \fi
}%
\providecommand \natexlab [1]{#1}%
\providecommand \enquote  [1]{``#1''}%
\providecommand \bibnamefont  [1]{#1}%
\providecommand \bibfnamefont [1]{#1}%
\providecommand \citenamefont [1]{#1}%
\providecommand \href@noop [0]{\@secondoftwo}%
\providecommand \href [0]{\begingroup \@sanitize@url \@href}%
\providecommand \@href[1]{\@@startlink{#1}\@@href}%
\providecommand \@@href[1]{\endgroup#1\@@endlink}%
\providecommand \@sanitize@url [0]{\catcode `\\12\catcode `\$12\catcode
  `\&12\catcode `\#12\catcode `\^12\catcode `\_12\catcode `\%12\relax}%
\providecommand \@@startlink[1]{}%
\providecommand \@@endlink[0]{}%
\providecommand \url  [0]{\begingroup\@sanitize@url \@url }%
\providecommand \@url [1]{\endgroup\@href {#1}{\urlprefix }}%
\providecommand \urlprefix  [0]{URL }%
\providecommand \Eprint [0]{\href }%
\providecommand \doibase [0]{http://dx.doi.org/}%
\providecommand \selectlanguage [0]{\@gobble}%
\providecommand \bibinfo  [0]{\@secondoftwo}%
\providecommand \bibfield  [0]{\@secondoftwo}%
\providecommand \translation [1]{[#1]}%
\providecommand \BibitemOpen [0]{}%
\providecommand \bibitemStop [0]{}%
\providecommand \bibitemNoStop [0]{.\EOS\space}%
\providecommand \EOS [0]{\spacefactor3000\relax}%
\providecommand \BibitemShut  [1]{\csname bibitem#1\endcsname}%
\let\auto@bib@innerbib\@empty
\bibitem [{\citenamefont {Hastings}(1970)}]{10.1093/biomet/57.1.97}%
  \BibitemOpen
  \bibfield  {author} {\bibinfo {author} {\bibfnamefont {W.~K.}\ \bibnamefont
  {Hastings}},\ }\href {\doibase 10.1093/biomet/57.1.97} {\bibfield  {journal}
  {\bibinfo  {journal} {Biometrika}\ }\textbf {\bibinfo {volume} {57}},\
  \bibinfo {pages} {97} (\bibinfo {year} {1970})}\BibitemShut {NoStop}%
\bibitem [{\citenamefont {{Geman}}\ and\ \citenamefont
  {{Geman}}(1984)}]{4767596}%
  \BibitemOpen
  \bibfield  {author} {\bibinfo {author} {\bibfnamefont {S.}~\bibnamefont
  {{Geman}}}\ and\ \bibinfo {author} {\bibfnamefont {D.}~\bibnamefont
  {{Geman}}},\ }\href@noop {} {\bibfield  {journal} {\bibinfo  {journal} {IEEE
  Transactions on Pattern Analysis and Machine Intelligence}\ }\textbf
  {\bibinfo {volume} {PAMI-6}},\ \bibinfo {pages} {721} (\bibinfo {year}
  {1984})}\BibitemShut {NoStop}%
\bibitem [{\citenamefont {Casella}\ \emph {et~al.}(2004)\citenamefont
  {Casella}, \citenamefont {Robert},\ and\ \citenamefont {Wells}}]{casella}%
  \BibitemOpen
  \bibfield  {author} {\bibinfo {author} {\bibfnamefont {G.}~\bibnamefont
  {Casella}}, \bibinfo {author} {\bibfnamefont {C.~P.}\ \bibnamefont {Robert}},
  \ and\ \bibinfo {author} {\bibfnamefont {M.~T.}\ \bibnamefont {Wells}},\
  }\enquote {\bibinfo {title} {Generalized accept-reject sampling schemes},}\
  in\ \href {\doibase 10.1214/lnms/1196285403} {\emph {\bibinfo {booktitle} {A
  Festschrift for Herman Rubin}}},\ \bibinfo {series} {Lecture Notes--Monograph
  Series}, Vol.\ \bibinfo {volume} {Volume 45},\ \bibinfo {editor} {edited by\
  \bibinfo {editor} {\bibfnamefont {A.}~\bibnamefont {DasGupta}}}\ (\bibinfo
  {publisher} {Institute of Mathematical Statistics},\ \bibinfo {address}
  {Beachwood, Ohio, USA},\ \bibinfo {year} {2004})\ pp.\ \bibinfo {pages}
  {342--347}\BibitemShut {NoStop}%
\bibitem [{\citenamefont {Martino}\ and\ \citenamefont
  {Míguez}(2010)}]{MARTINO20102981}%
  \BibitemOpen
  \bibfield  {author} {\bibinfo {author} {\bibfnamefont {L.}~\bibnamefont
  {Martino}}\ and\ \bibinfo {author} {\bibfnamefont {J.}~\bibnamefont
  {Míguez}},\ }\href {\doibase https://doi.org/10.1016/j.sigpro.2010.04.025}
  {\bibfield  {journal} {\bibinfo  {journal} {Signal Processing}\ }\textbf
  {\bibinfo {volume} {90}},\ \bibinfo {pages} {2981 } (\bibinfo {year}
  {2010})}\BibitemShut {NoStop}%
\bibitem [{\citenamefont {Bishop}\ and\ \citenamefont
  {Nasrabadi}(2007)}]{Bishop2007PatternRA}%
  \BibitemOpen
  \bibfield  {author} {\bibinfo {author} {\bibfnamefont {C.~M.}\ \bibnamefont
  {Bishop}}\ and\ \bibinfo {author} {\bibfnamefont {N.~M.}\ \bibnamefont
  {Nasrabadi}},\ }\href@noop {} {\bibfield  {journal} {\bibinfo  {journal} {J.
  Electronic Imaging}\ }\textbf {\bibinfo {volume} {16}},\ \bibinfo {pages}
  {049901} (\bibinfo {year} {2007})}\BibitemShut {NoStop}%
\bibitem [{\citenamefont {von Neumann}(1951)}]{vonNeumann1951}%
  \BibitemOpen
  \bibfield  {author} {\bibinfo {author} {\bibfnamefont {J.}~\bibnamefont {von
  Neumann}},\ }in\ \href@noop {} {\emph {\bibinfo {booktitle} {Monte Carlo
  Method}}},\ \bibinfo {series} {National Bureau of Standards Applied
  Mathematics Series}, Vol.~\bibinfo {volume} {12},\ \bibinfo {editor} {edited
  by\ \bibinfo {editor} {\bibfnamefont {A.~S.}\ \bibnamefont {Householder}},
  \bibinfo {editor} {\bibfnamefont {G.~E.}\ \bibnamefont {Forsythe}}, \ and\
  \bibinfo {editor} {\bibfnamefont {H.~H.}\ \bibnamefont {Germond}}}\ (\bibinfo
   {publisher} {US Government Printing Office},\ \bibinfo {address}
  {Washington, DC},\ \bibinfo {year} {1951})\ Chap.~\bibinfo {chapter} {13},
  pp.\ \bibinfo {pages} {36--38}\BibitemShut {NoStop}%
\bibitem [{\citenamefont {Robert}\ and\ \citenamefont
  {Casella}(2004)}]{Robert2004}%
  \BibitemOpen
  \bibfield  {author} {\bibinfo {author} {\bibfnamefont {C.~P.}\ \bibnamefont
  {Robert}}\ and\ \bibinfo {author} {\bibfnamefont {G.}~\bibnamefont
  {Casella}},\ }\href {\doibase 10.1007/978-1-4757-4145-2} {\emph {\bibinfo
  {title} {Monte Carlo Statistical Methods}}}\ (\bibinfo  {publisher} {Springer
  New York},\ \bibinfo {year} {2004})\BibitemShut {NoStop}%
\bibitem [{\citenamefont {M{\"u}ller}\ \emph {et~al.}(2018)\citenamefont
  {M{\"u}ller}, \citenamefont {McWilliams}, \citenamefont {Rousselle},
  \citenamefont {Gross},\ and\ \citenamefont {Nov{\'a}k}}]{Mller2018NeuralIS}%
  \BibitemOpen
  \bibfield  {author} {\bibinfo {author} {\bibfnamefont {T.}~\bibnamefont
  {M{\"u}ller}}, \bibinfo {author} {\bibfnamefont {B.}~\bibnamefont
  {McWilliams}}, \bibinfo {author} {\bibfnamefont {F.}~\bibnamefont
  {Rousselle}}, \bibinfo {author} {\bibfnamefont {M.}~\bibnamefont {Gross}}, \
  and\ \bibinfo {author} {\bibfnamefont {J.}~\bibnamefont {Nov{\'a}k}},\
  }\href@noop {} {\bibfield  {journal} {\bibinfo  {journal} {ACM Trans.
  Graph.}\ }\textbf {\bibinfo {volume} {38}},\ \bibinfo {pages} {145:1}
  (\bibinfo {year} {2018})}\BibitemShut {NoStop}%
\bibitem [{\citenamefont {Kahn}\ and\ \citenamefont
  {Marshall}(1953)}]{10.2307/166789}%
  \BibitemOpen
  \bibfield  {author} {\bibinfo {author} {\bibfnamefont {H.}~\bibnamefont
  {Kahn}}\ and\ \bibinfo {author} {\bibfnamefont {A.~W.}\ \bibnamefont
  {Marshall}},\ }\href {http://www.jstor.org/stable/166789} {\bibfield
  {journal} {\bibinfo  {journal} {Journal of the Operations Research Society of
  America}\ }\textbf {\bibinfo {volume} {1}},\ \bibinfo {pages} {263} (\bibinfo
  {year} {1953})}\BibitemShut {NoStop}%
\bibitem [{\citenamefont {Bothmann}\ \emph {et~al.}(2020)\citenamefont
  {Bothmann}, \citenamefont {Janßen}, \citenamefont {Knobbe}, \citenamefont
  {Schmale},\ and\ \citenamefont {Schumann}}]{Bothmann:2020ywa}%
  \BibitemOpen
  \bibfield  {author} {\bibinfo {author} {\bibfnamefont {E.}~\bibnamefont
  {Bothmann}}, \bibinfo {author} {\bibfnamefont {T.}~\bibnamefont {Janßen}},
  \bibinfo {author} {\bibfnamefont {M.}~\bibnamefont {Knobbe}}, \bibinfo
  {author} {\bibfnamefont {T.}~\bibnamefont {Schmale}}, \ and\ \bibinfo
  {author} {\bibfnamefont {S.}~\bibnamefont {Schumann}},\ }\href {\doibase
  10.21468/SciPostPhys.8.4.069} {\bibfield  {journal} {\bibinfo  {journal}
  {SciPost Phys.}\ }\textbf {\bibinfo {volume} {8}},\ \bibinfo {pages} {069}
  (\bibinfo {year} {2020})},\ \Eprint {http://arxiv.org/abs/2001.05478}
  {arXiv:2001.05478 [hep-ph]} \BibitemShut {NoStop}%
\bibitem [{\citenamefont {Gao}\ \emph {et~al.}(2020)\citenamefont {Gao},
  \citenamefont {H\"oche}, \citenamefont {Isaacson}, \citenamefont {Krause},\
  and\ \citenamefont {Schulz}}]{PhysRevD.101.076002}%
  \BibitemOpen
  \bibfield  {author} {\bibinfo {author} {\bibfnamefont {C.}~\bibnamefont
  {Gao}}, \bibinfo {author} {\bibfnamefont {S.}~\bibnamefont {H\"oche}},
  \bibinfo {author} {\bibfnamefont {J.}~\bibnamefont {Isaacson}}, \bibinfo
  {author} {\bibfnamefont {C.}~\bibnamefont {Krause}}, \ and\ \bibinfo {author}
  {\bibfnamefont {H.}~\bibnamefont {Schulz}},\ }\href {\doibase
  10.1103/PhysRevD.101.076002} {\bibfield  {journal} {\bibinfo  {journal}
  {Phys. Rev. D}\ }\textbf {\bibinfo {volume} {101}},\ \bibinfo {pages}
  {076002} (\bibinfo {year} {2020})}\BibitemShut {NoStop}%
\bibitem [{\citenamefont {Alvarez-Ruso}\ \emph {et~al.}(2018)\citenamefont
  {Alvarez-Ruso} \emph {et~al.}}]{Alvarez-Ruso:2017oui}%
  \BibitemOpen
  \bibfield  {author} {\bibinfo {author} {\bibfnamefont {L.}~\bibnamefont
  {Alvarez-Ruso}} \emph {et~al.} (\bibinfo {collaboration} {NuSTEC}),\ }\href
  {\doibase 10.1016/j.ppnp.2018.01.006} {\bibfield  {journal} {\bibinfo
  {journal} {Prog. Part. Nucl. Phys.}\ }\textbf {\bibinfo {volume} {100}},\
  \bibinfo {pages} {1} (\bibinfo {year} {2018})},\ \Eprint
  {http://arxiv.org/abs/1706.03621} {arXiv:1706.03621 [hep-ph]} \BibitemShut
  {NoStop}%
\bibitem [{\citenamefont {Abe}\ \emph {et~al.}(2020)\citenamefont {Abe} \emph
  {et~al.}}]{Abe:2019vii}%
  \BibitemOpen
  \bibfield  {author} {\bibinfo {author} {\bibfnamefont {K.}~\bibnamefont
  {Abe}} \emph {et~al.} (\bibinfo {collaboration} {T2K}),\ }\href {\doibase
  10.1038/s41586-020-2177-0} {\bibfield  {journal} {\bibinfo  {journal}
  {Nature}\ }\textbf {\bibinfo {volume} {580}},\ \bibinfo {pages} {339}
  (\bibinfo {year} {2020})},\ \Eprint {http://arxiv.org/abs/1910.03887}
  {1910.03887} \BibitemShut {NoStop}%
\bibitem [{\citenamefont {Acero}\ \emph {et~al.}(2019)\citenamefont {Acero}
  \emph {et~al.}}]{Acero:2019ksn}%
  \BibitemOpen
  \bibfield  {author} {\bibinfo {author} {\bibfnamefont {M.}~\bibnamefont
  {Acero}} \emph {et~al.} (\bibinfo {collaboration} {NOvA}),\ }\href {\doibase
  10.1103/PhysRevLett.123.151803} {\bibfield  {journal} {\bibinfo  {journal}
  {Phys. Rev. Lett.}\ }\textbf {\bibinfo {volume} {123}},\ \bibinfo {pages}
  {151803} (\bibinfo {year} {2019})},\ \Eprint
  {http://arxiv.org/abs/1906.04907} {arXiv:1906.04907 [hep-ex]} \BibitemShut
  {NoStop}%
\bibitem [{\citenamefont {Smith}\ and\ \citenamefont
  {Moniz}(1972)}]{Smith:1972xh}%
  \BibitemOpen
  \bibfield  {author} {\bibinfo {author} {\bibfnamefont {R.}~\bibnamefont
  {Smith}}\ and\ \bibinfo {author} {\bibfnamefont {E.}~\bibnamefont {Moniz}},\
  }\href {\doibase 10.1016/0550-3213(75)90612-4} {\bibfield  {journal}
  {\bibinfo  {journal} {Nucl. Phys. B}\ }\textbf {\bibinfo {volume} {43}},\
  \bibinfo {pages} {605} (\bibinfo {year} {1972})},\ \bibinfo {note} {[Erratum:
  Nucl.Phys.B 101, 547 (1975)]}\BibitemShut {NoStop}%
\bibitem [{\citenamefont {Papamakarios}\ \emph {et~al.}(2019)\citenamefont
  {Papamakarios}, \citenamefont {Nalisnick}, \citenamefont {Rezende},
  \citenamefont {Mohamed},\ and\ \citenamefont
  {Lakshminarayanan}}]{Papamakarios2019NormalizingFF}%
  \BibitemOpen
  \bibfield  {author} {\bibinfo {author} {\bibfnamefont {G.}~\bibnamefont
  {Papamakarios}}, \bibinfo {author} {\bibfnamefont {E.~T.}\ \bibnamefont
  {Nalisnick}}, \bibinfo {author} {\bibfnamefont {D.~J.}\ \bibnamefont
  {Rezende}}, \bibinfo {author} {\bibfnamefont {S.}~\bibnamefont {Mohamed}}, \
  and\ \bibinfo {author} {\bibfnamefont {B.}~\bibnamefont {Lakshminarayanan}},\
  }\href@noop {} {\bibfield  {journal} {\bibinfo  {journal} {ArXiv}\ }\textbf
  {\bibinfo {volume} {abs/1912.02762}} (\bibinfo {year} {2019})}\BibitemShut
  {NoStop}%
\bibitem [{\citenamefont {Durkan}\ \emph
  {et~al.}(2019{\natexlab{a}})\citenamefont {Durkan}, \citenamefont {Bekasov},
  \citenamefont {Murray},\ and\ \citenamefont
  {Papamakarios}}]{Durkan2019NeuralSF}%
  \BibitemOpen
  \bibfield  {author} {\bibinfo {author} {\bibfnamefont {C.}~\bibnamefont
  {Durkan}}, \bibinfo {author} {\bibfnamefont {A.}~\bibnamefont {Bekasov}},
  \bibinfo {author} {\bibfnamefont {I.}~\bibnamefont {Murray}}, \ and\ \bibinfo
  {author} {\bibfnamefont {G.}~\bibnamefont {Papamakarios}},\ }in\ \href
  {http://papers.nips.cc/paper/8969-neural-spline-flows.pdf} {\emph {\bibinfo
  {booktitle} {Advances in Neural Information Processing Systems 32}}}\
  (\bibinfo  {publisher} {Curran Associates, Inc.},\ \bibinfo {year} {2019})\
  pp.\ \bibinfo {pages} {7511--7522}\BibitemShut {NoStop}%
\bibitem [{\citenamefont {Rezende}\ and\ \citenamefont
  {Mohamed}(2015)}]{Rezende2015VariationalIW}%
  \BibitemOpen
  \bibfield  {author} {\bibinfo {author} {\bibfnamefont {D.~J.}\ \bibnamefont
  {Rezende}}\ and\ \bibinfo {author} {\bibfnamefont {S.}~\bibnamefont
  {Mohamed}},\ }\href@noop {} {\bibfield  {journal} {\bibinfo  {journal}
  {ArXiv}\ }\textbf {\bibinfo {volume} {abs/1505.05770}} (\bibinfo {year}
  {2015})}\BibitemShut {NoStop}%
\bibitem [{\citenamefont {Germain}\ \emph {et~al.}(2015)\citenamefont
  {Germain}, \citenamefont {Gregor}, \citenamefont {Murray},\ and\
  \citenamefont {Larochelle}}]{Germain2015MADEMA}%
  \BibitemOpen
  \bibfield  {author} {\bibinfo {author} {\bibfnamefont {M.}~\bibnamefont
  {Germain}}, \bibinfo {author} {\bibfnamefont {K.}~\bibnamefont {Gregor}},
  \bibinfo {author} {\bibfnamefont {I.}~\bibnamefont {Murray}}, \ and\ \bibinfo
  {author} {\bibfnamefont {H.}~\bibnamefont {Larochelle}},\ }in\ \href
  {http://proceedings.mlr.press/v37/germain15.html} {\emph {\bibinfo
  {booktitle} {Proceedings of the 32nd International Conference on Machine
  Learning}}},\ \bibinfo {series} {Proceedings of Machine Learning Research},
  Vol.~\bibinfo {volume} {37},\ \bibinfo {editor} {edited by\ \bibinfo {editor}
  {\bibfnamefont {F.}~\bibnamefont {Bach}}\ and\ \bibinfo {editor}
  {\bibfnamefont {D.}~\bibnamefont {Blei}}}\ (\bibinfo  {publisher} {PMLR},\
  \bibinfo {address} {Lille, France},\ \bibinfo {year} {2015})\ pp.\ \bibinfo
  {pages} {881--889}\BibitemShut {NoStop}%
\bibitem [{\citenamefont {Papamakarios}\ \emph {et~al.}(2017)\citenamefont
  {Papamakarios}, \citenamefont {Pavlakou},\ and\ \citenamefont
  {Murray}}]{Papamakarios2017MaskedAF}%
  \BibitemOpen
  \bibfield  {author} {\bibinfo {author} {\bibfnamefont {G.}~\bibnamefont
  {Papamakarios}}, \bibinfo {author} {\bibfnamefont {T.}~\bibnamefont
  {Pavlakou}}, \ and\ \bibinfo {author} {\bibfnamefont {I.}~\bibnamefont
  {Murray}},\ }in\ \href
  {http://papers.nips.cc/paper/6828-masked-autoregressive-flow-for-density-estimation.pdf}
  {\emph {\bibinfo {booktitle} {Advances in Neural Information Processing
  Systems 30}}},\ \bibinfo {editor} {edited by\ \bibinfo {editor}
  {\bibfnamefont {I.}~\bibnamefont {Guyon}}, \bibinfo {editor} {\bibfnamefont
  {U.~V.}\ \bibnamefont {Luxburg}}, \bibinfo {editor} {\bibfnamefont
  {S.}~\bibnamefont {Bengio}}, \bibinfo {editor} {\bibfnamefont
  {H.}~\bibnamefont {Wallach}}, \bibinfo {editor} {\bibfnamefont
  {R.}~\bibnamefont {Fergus}}, \bibinfo {editor} {\bibfnamefont
  {S.}~\bibnamefont {Vishwanathan}}, \ and\ \bibinfo {editor} {\bibfnamefont
  {R.}~\bibnamefont {Garnett}}}\ (\bibinfo  {publisher} {Curran Associates,
  Inc.},\ \bibinfo {year} {2017})\ pp.\ \bibinfo {pages}
  {2338--2347}\BibitemShut {NoStop}%
\bibitem [{\citenamefont {Huang}\ \emph {et~al.}(2018)\citenamefont {Huang},
  \citenamefont {Krueger}, \citenamefont {Lacoste},\ and\ \citenamefont
  {Courville}}]{Huang2018NeuralAF}%
  \BibitemOpen
  \bibfield  {author} {\bibinfo {author} {\bibfnamefont {C.-W.}\ \bibnamefont
  {Huang}}, \bibinfo {author} {\bibfnamefont {D.}~\bibnamefont {Krueger}},
  \bibinfo {author} {\bibfnamefont {A.}~\bibnamefont {Lacoste}}, \ and\
  \bibinfo {author} {\bibfnamefont {A.~C.}\ \bibnamefont {Courville}},\
  }\href@noop {} {\bibfield  {journal} {\bibinfo  {journal} {ArXiv}\ }\textbf
  {\bibinfo {volume} {abs/1804.00779}} (\bibinfo {year} {2018})}\BibitemShut
  {NoStop}%
\bibitem [{\citenamefont {Cao}\ \emph {et~al.}(2019)\citenamefont {Cao},
  \citenamefont {Titov},\ and\ \citenamefont {Aziz}}]{Cao2019BlockNA}%
  \BibitemOpen
  \bibfield  {author} {\bibinfo {author} {\bibfnamefont {N.~D.}\ \bibnamefont
  {Cao}}, \bibinfo {author} {\bibfnamefont {I.}~\bibnamefont {Titov}}, \ and\
  \bibinfo {author} {\bibfnamefont {W.}~\bibnamefont {Aziz}},\ }\href@noop {}
  {\enquote {\bibinfo {title} {Block neural autoregressive flow},}\ } (\bibinfo
  {year} {2019}),\ \Eprint {http://arxiv.org/abs/1904.04676} {arXiv:1904.04676
  [stat.ML]} \BibitemShut {NoStop}%
\bibitem [{\citenamefont {Wehenkel}\ and\ \citenamefont
  {Louppe}(2019)}]{Wehenkel2019UnconstrainedMN}%
  \BibitemOpen
  \bibfield  {author} {\bibinfo {author} {\bibfnamefont {A.}~\bibnamefont
  {Wehenkel}}\ and\ \bibinfo {author} {\bibfnamefont {G.}~\bibnamefont
  {Louppe}},\ }in\ \href
  {http://papers.nips.cc/paper/8433-unconstrained-monotonic-neural-networks.pdf}
  {\emph {\bibinfo {booktitle} {Advances in Neural Information Processing
  Systems 32}}}\ (\bibinfo  {publisher} {Curran Associates, Inc.},\ \bibinfo
  {year} {2019})\ pp.\ \bibinfo {pages} {1545--1555}\BibitemShut {NoStop}%
\bibitem [{\citenamefont {Jaini}\ \emph {et~al.}(2019)\citenamefont {Jaini},
  \citenamefont {Selby},\ and\ \citenamefont {Yu}}]{Jaini2019SumofSquaresPF}%
  \BibitemOpen
  \bibfield  {author} {\bibinfo {author} {\bibfnamefont {P.}~\bibnamefont
  {Jaini}}, \bibinfo {author} {\bibfnamefont {K.~A.}\ \bibnamefont {Selby}}, \
  and\ \bibinfo {author} {\bibfnamefont {Y.}~\bibnamefont {Yu}},\ }in\ \href
  {http://proceedings.mlr.press/v97/jaini19a.html} {\emph {\bibinfo {booktitle}
  {Proceedings of the 36th International Conference on Machine Learning}}},\
  \bibinfo {series} {Proceedings of Machine Learning Research}, Vol.~\bibinfo
  {volume} {97},\ \bibinfo {editor} {edited by\ \bibinfo {editor}
  {\bibfnamefont {K.}~\bibnamefont {Chaudhuri}}\ and\ \bibinfo {editor}
  {\bibfnamefont {R.}~\bibnamefont {Salakhutdinov}}}\ (\bibinfo  {publisher}
  {PMLR},\ \bibinfo {address} {Long Beach, California, USA},\ \bibinfo {year}
  {2019})\ pp.\ \bibinfo {pages} {3009--3018}\BibitemShut {NoStop}%
\bibitem [{\citenamefont {Durkan}\ \emph
  {et~al.}(2019{\natexlab{b}})\citenamefont {Durkan}, \citenamefont {Bekasov},
  \citenamefont {Murray},\ and\ \citenamefont
  {Papamakarios}}]{Durkan2019CubicSplineF}%
  \BibitemOpen
  \bibfield  {author} {\bibinfo {author} {\bibfnamefont {C.}~\bibnamefont
  {Durkan}}, \bibinfo {author} {\bibfnamefont {A.}~\bibnamefont {Bekasov}},
  \bibinfo {author} {\bibfnamefont {I.}~\bibnamefont {Murray}}, \ and\ \bibinfo
  {author} {\bibfnamefont {G.}~\bibnamefont {Papamakarios}},\ }\href@noop {}
  {\bibfield  {journal} {\bibinfo  {journal} {ArXiv}\ }\textbf {\bibinfo
  {volume} {abs/1906.02145}} (\bibinfo {year}
  {2019}{\natexlab{b}})}\BibitemShut {NoStop}%
\bibitem [{\citenamefont {Gregory}\ and\ \citenamefont
  {Delbourgo}(1982)}]{Gregory1982PiecewiseRQ}%
  \BibitemOpen
  \bibfield  {author} {\bibinfo {author} {\bibfnamefont {J.~A.}\ \bibnamefont
  {Gregory}}\ and\ \bibinfo {author} {\bibfnamefont {R.}~\bibnamefont
  {Delbourgo}},\ }\href {\doibase 10.1093/imanum/2.2.123} {\bibfield  {journal}
  {\bibinfo  {journal} {IMA Journal of Numerical Analysis}\ }\textbf {\bibinfo
  {volume} {2}},\ \bibinfo {pages} {123} (\bibinfo {year} {1982})},\ \Eprint
  {http://arxiv.org/abs/https://academic.oup.com/imajna/article-pdf/2/2/123/2267745/2-2-123.pdf}
  {https://academic.oup.com/imajna/article-pdf/2/2/123/2267745/2-2-123.pdf}
  \BibitemShut {NoStop}%
\bibitem [{\citenamefont {Kullback}\ and\ \citenamefont
  {Leibler}(1951)}]{kullback1951}%
  \BibitemOpen
  \bibfield  {author} {\bibinfo {author} {\bibfnamefont {S.}~\bibnamefont
  {Kullback}}\ and\ \bibinfo {author} {\bibfnamefont {R.~A.}\ \bibnamefont
  {Leibler}},\ }\href {\doibase 10.1214/aoms/1177729694} {\bibfield  {journal}
  {\bibinfo  {journal} {Ann. Math. Statist.}\ }\textbf {\bibinfo {volume}
  {22}},\ \bibinfo {pages} {79} (\bibinfo {year} {1951})}\BibitemShut {NoStop}%
\bibitem [{\citenamefont {Kingma}\ and\ \citenamefont
  {Ba}(2014)}]{Kingma2014AdamAM}%
  \BibitemOpen
  \bibfield  {author} {\bibinfo {author} {\bibfnamefont {D.~P.}\ \bibnamefont
  {Kingma}}\ and\ \bibinfo {author} {\bibfnamefont {J.}~\bibnamefont {Ba}},\
  }\href@noop {} {\bibfield  {journal} {\bibinfo  {journal} {CoRR}\ }\textbf
  {\bibinfo {volume} {abs/1412.6980}} (\bibinfo {year} {2014})}\BibitemShut
  {NoStop}%
\bibitem [{\citenamefont {Liu}(1996)}]{Liu1996}%
  \BibitemOpen
  \bibfield  {author} {\bibinfo {author} {\bibfnamefont {J.~S.}\ \bibnamefont
  {Liu}},\ }\href {\doibase 10.1007/bf00162521} {\bibfield  {journal} {\bibinfo
   {journal} {Statistics and Computing}\ }\textbf {\bibinfo {volume} {6}},\
  \bibinfo {pages} {113} (\bibinfo {year} {1996})}\BibitemShut {NoStop}%
\bibitem [{Note1()}]{Note1}%
  \BibitemOpen
  \bibinfo {note} {The concept of the Asimov data is being a representative
  date set, inspired by the short story Franchise by Isaac Asimov, in which the
  entire electorate is replaced by selecting the most representative
  voter.}\BibitemShut {Stop}%
\end{thebibliography}%

\appendix

\section{Model complexity discussion}
\label{Sec:appendixComplexity}

We can see in Tab.~\ref{Tab:Performance} that although the number of rejection sampling cycles is remarkably lower for the NSF proposal, the time for obtaining one million samples is fairly similar. This is because the time for a cycle is the sum of the time in generating and evaluating the proposal plus the time it takes to evaluate $\probp{\sampleX}$, in this case the CCQE cross-section. On average, for one rejection cycle of this experiment, generating+sampling for the NSF a batch of $300\,000$~samples takes $0.31$~seconds, for a uniform proposal it only takes $0.00079$~seconds, and evaluating $300\,000$ samples for the CCQE cross-section model takes $0.0021$~seconds. When summing up the time for the selected proposal plus the time of the model, we can see that the NSF takes a large fraction of the computational time (more than hundred times more than evaluating the cross-section model), so overall a cycle for the NSF proposal takes over a hundred cycles of uniform proposal. This is because the CCQE cross-section model we have chosen for this study (Sec.~\ref{Sec:CCQEXsect}) is relatively simple, especially compared to other applications in HEP. Additionally we are still relatively low in the number of dimensions. As a thought experiment, we considered different cases for a similar set up: sampling ten million samples using batches of $300\,000$ in each cycle, where the time of evaluating the model $\probp{\sampleX}$ is increased by a factor of $t_\mathrm{increase}$.

\begin{table}[!thp]
    \centering
    \caption{Model complexity time comparison table for both NSF and uniform proposals to generate ten million samples via rejection sampling. Choosing $k_Q$ as one of the below quantiles, we increase the time to evaluate the model $\probp{\sampleX}$ for each rejection cycle by a factor $t_\mathrm{increase}$. $t_\mathrm{NSF}$ ($t_\mathrm{Unif}$) is the time it takes to obtain ten million accepted samples, in seconds, for the NSF (uniform) proposal. $t_\mathrm{Unif}/t_\mathrm{NSF}$ is the ratio of the time between both proposals, showing that for more complex models than the CCQE cross-section of this paper, NSF outperforms a uniform proposal by orders of magnitude.}
    \label{tab:increase}
    \begin{tabular}{llrrr}
        Quantile & $t_\mathrm{increase}$ &  $t_\mathrm{NSF}$ &  $t_\mathrm{Unif}$ &  $t_\mathrm{Unif}/t_\mathrm{NSF}$ \\
        \hline
        1.0000 & 1    &        2017.26 &         471.64 &         0.23 \\
               & 10   &        2141.10 &        3567.01 &         1.67 \\
               & 100  &        3379.54 &       34520.76 &        10.21 \\
               & 1000 &       15763.94 &      344058.17 &        21.83 \\
        \hline
        0.9999 & 1    &          63.95 &         123.19 &         1.93 \\
               & 10   &          67.87 &         928.73 &        13.68 \\
               & 100  &         107.01 &        8984.12 &        83.96 \\
               & 1000 &         498.39 &       89538.01 &       179.65 \\
       \hline
        0.9990 & 1    &          28.87 &          35.36 &         1.22 \\
               & 10   &          30.64 &         266.54 &         8.70 \\
               & 100  &          48.31 &        2578.40 &        53.37 \\
               & 1000 &         225.00 &       25696.93 &       114.21 \\
    \end{tabular}
\end{table}

Tab.~\ref{tab:increase} shows the results for obtaining one million samples under these circumstances for the quantiles $\kQ{1.0}$,  $\kQ{0.9999}$ and $\kQ{0.9990}$ with the same acceptance probability as in Tab.~\ref{Tab:Performance}, consider different hypothetical models with factors $t_\mathrm{increase}\in\{1,10,100,1000\}$ compared to the CCQE cross-section. We show the time (in seconds) it takes to generate these ten milllion samples for the NSF, $t_\mathrm{NSF}$, and for the uniform proposal, $t_\mathrm{Unif}$, while also computing their ratio $t_\mathrm{Unif}/t_\mathrm{NSF}$. For models  of $\probp{\sampleX}$ with larger computationally complexity, we can see that even for the maximum quantile $\kQ{1.0}$ NSF is faster to obtain ten million accepted samples than using a uniform proposal, gaining a whole order of magnitude in time when the model is at least hundred times more complex. Choosing quantiles smaller than 1.0, for models which take a thousand times longer to compute compared to the CCQE cross-section one, the NSF proposal generates these ten million samples over a hundred times faster than the uniform distribution. Even for simpler models with only hundred more computation complexity, the gain of using the NSF proposal is noticeable. 

Regarding the number of parameters used by the NSF with increasing number of dimensions, when maintaining the same architecture for the NSF, the number scales linearly with the dimension of the data. However, when considering higher dimensions, to enrich further the flexibility of the family of densities produced by the NSF to account for more complex correlations and dependencies between each dimension, the architecture might have to change, adding an increase larger than linear for the number of parameters.
\end{document}